\RequirePackage{fix-cm}
\documentclass[twocolumn]{svjour3}          
\smartqed  
\usepackage{graphicx}
\usepackage{epsfig}
\usepackage{amsmath}
\usepackage{rotating}
\usepackage{fancyvrb}
\usepackage{algorithm}
\usepackage{algorithmic}
\usepackage{balance}  
\usepackage{array}

\journalname{}
\begin{document}

\title{Revisiting Aggregation for Data Intensive Applications:\newline A Performance Study
}


\author{Jian Wen \and
        Vinayak R. Borkar \and
        Michael J. Carey \and
        Vassilis J. Tsotras
}


\institute{Jian Wen, Vassilis J. Tsotras \at
              Univ. of California, Riverside \\
              \email{wenj@cs.ucr.edu, tsotras@cs.ucr.edu}           
           \and
           Vinayak R. Borkar, Michael J. Carey \at
           Univ. of California, Irvine\\
           \email{vborkar@ics.uci.edu, mjcarey@ics.uci.edu}
}

\date{Received: date / Accepted: date}

\maketitle

\begin{abstract}
Aggregation has been an important operation since the early days of
relational databases.
Today's Big Data applications bring further challenges when processing
aggregation queries, demanding
adaptive aggregation algorithms that can process large volumes of data
relative to a potentially limited
memory budget (especially in multiuser settings).
Despite its importance, the design and evaluation of aggregation
algorithms has not received the
same attention that other basic operators, such as joins, have
received in the literature.
As a result, when considering which aggregation algorithm(s) to
implement in a new parallel Big Data
processing platform (AsterixDB), we faced a lack of ``off the shelf''
answers that we could simply read
about and then implement based on prior performance studies.

In this paper we revisit the engineering of efficient local aggregation
algorithms for use in Big Data platforms.
We discuss the salient implementation details of several candidate
algorithms and present an in-depth
experimental performance study to guide future Big Data engine developers.
We show that the efficient implementation of the aggregation operator
for a Big Data platform is non-trivial
and that many factors, including memory usage, spilling strategy, and
I/O and CPU cost, should be considered.
Further, we introduce precise cost models that can help in choosing an
appropriate algorithm based on input
parameters including memory budget, grouping key cardinality, and data skew.
\end{abstract}
%
%
%

\section{Introduction}
\label{sec:intro}

Aggregation has always been a very important operation in database processing. For example, all 22 queries in the TPC-H Benchmark \cite{bibd:url///tpc-h-benchmark} contain aggregation. It is also a key operation for data preprocessing and query processing in data intensive applications, such as machine learning on large volume data \cite{DBLP:conf/nips/ChuKLYBNO06}, and web-related data processing like web-log and page ranking \cite{DBLP:conf/sigmod/PavloPRADMS09}, etc. In the big data scenario where data is spread over a distributed environment, like Hadoop and many popular distributed relational databases, aggregation is typically processed in a map-combine-reduce fashion. Such a strategy first obtains the local aggregation results, which are then merged to get the global aggregation results. Hence, the efficiency of the local aggregation algorithm is a key factor for the global aggregation performance.

In our effort to support the aggregation operation in our next generation parallel data processing platform AsterixDB \cite{DBLP:journals/pvldb/AlsubaieeAABBBCGHKLOPVW12n}, we noticed two new challenges that big data applications impose on local aggregation algorithms: first, if the input data is huge and the aggregation is group-based (like the ``group-by'' in SQL, where each unique group will have a record in the result set), the aggregation result may not fit in main memory; second, in order to allow multiple operations being processed simultaneously, an aggregation operation should work within a strict memory budget provided by the platform. 

Implementing an aggregation operation to address these challenges is not trivial. Aggregation has not attracted as much attention as other operations like joins, probably due to its simpler computational logic. 
Several aggregation algorithms proposed in literature decades ago, like pre-sorting the input data \cite{bibd:techreport//epsteine1979/techniques-for-processing}, or using hashing \cite{DBLP:conf/sigmod/ShatdalN95}, have not been fully studied with respect to their performance for very large datasets or datasets with different distribution properties. While some join processing techniques \cite{DBLP:conf/vldb/GraefeBC98} can be adapted for aggregation queries, they are tuned for better join performance. All these existing algorithms lack for details on how to implement them using strictly bounded memory, and there is no study about which aggregation algorithm works better for which circumstances. To answer these questions we present in this paper a thorough study of single machine aggregation algorithms under the bounded memory and big data assumptions.
Our contributions can be summarized as:

\begin{enumerate}

\item We present detailed implementation strategies for six aggregation algorithms: two are novel and four are based on extending existing  algorithms. All algorithms work within a strictly bounded memory budget, and they can easily adapt between in-memory and external processing.
\item We devise precise theoretical cost models for the algorithms' CPU and I/O behaviors. Based on input parameters, such models
can be used by a query optimizer to choose the right aggregation strategy.
\item We deploy all algorithms as operators on the Hyracks platform \cite{DBLP:conf/icde/BorkarCGOV11}, a flexible, extensible, partitioned-parallel platform for data-intensive computing, and evaluate their performance through extensive experimentation.
\end{enumerate}

Note that this paper is the first part of a two-part big data aggregation study; here we address the ``map'' phase with 
 extensive study of local aggregation algorithms. The result of this study provides a foundation for proper local aggregation algorithms as the component of a global aggregation strategy for the next, ``reduce'' phase, study.

The rest of the paper is organized as follows: Section~\ref{sec:rel} presents related research, while Section~\ref{sec:proenv} discusses the processing environment for our aggregation algorithms. Section~\ref{sec:algs} describes in detail all algorithms and Section~\ref{sec:theo} presents their theoretical performance analysis. The experimental evaluation results appear in Section~\ref{sec:eval}. Section~\ref{sec:opt} discusses the algorithm selection strategy in AsterixDB, and Section~\ref{sec:conclu} concludes the paper. In the Appendix we list the theoretical details of the basic component models used in our cost model analysis in Section~\ref{sec:theo}.

\section{Related Work}\label{sec:rel}

In our search for efficient local aggregation algorithms for AsterixDB, we noticed that aggregation has not drawn much attention in the study of efficient algorithms using tightly bounded memory. The well-known sort-based and hash-based aggregation algorithms discussed in \cite{bibd:techreport//epsteine1979/techniques-for-processing}, \cite{DBLP:journals/tods/BittonBDW83}, \cite{DBLP:books/daglib/0011128} and \cite{DBLP:books/daglib/0015084} provide straight-forward approaches to handle both in-memory and external aggregations, but these algorithms use sorting and hashing in a straight-forward way and there is space to further optimize the CPU and I/O cost. 
\cite{DBLP:journals/csur/Graefe93} discussed three approaches for aggregations that may not fit into memory, namely nested-loop, sort-based and hash-based. It suggests that the hash-based approach using hybrid-hash would be the choice when the input data can be greatly collapsed through aggregation. Our study of the hybrid-hash algorithm reveals that its hashing and partitioning strategy can be implemented in different ways, leading to different performance behaviors. These have not been discussed in the original paper, and precise cost models are also missing for the proper selection of aggregation algorithms under different configurations. 
\cite{DBLP:conf/vldb/GraefeBC98} presented optimizations for hybrid-hash-based algorithms, including dynamic destaging, partition tuning and many best-practice experiences from the experience of SQL Server implementation. However, this paper focuses more on optimization related to joins rather than aggregations.
\cite{DBLP:conf/sigmod/ShatdalN95} tried to address the problem of efficient parallel aggregation algorithms albeit for SQL, as we are doing for the AsterixDB project. For the local aggregation algorithm, they picked a variant of the hybrid-hash aggregation algorithm that shares its hashing space among all partitions. But no optimization has been done with other aggregation algorithms. 
More recently, \cite{DBLP:conf/vldb/CieslewiczR07} examined thread-level parallelism and proposed an adaptive aggregation algorithm optimized for cache locality by sampling and sharing the hash table in cache. However, in order to reveal the performance benefits from using the CPU cache, only in-memory aggregation algorithms were addressed. We think that for an external aggregation algorithm, it is important to address the I/O efficiency first, and then to optimize the CPU behavior for each in-memory run of the aggregation. 
\cite{DBLP:conf/damon/YeRV11} studied several in-memory aggregation algorithms for efficient thread-level parallelism and reducing cache contention. Similar to our proposed Pre-Partitioning algorithm, the PLAT algorithm in their paper partitions the input data based on their input order and fills up the per-thread hash table first. However, PLAT processes records in memory even after the hash table is full, based on the assumption that the input data can be fit into memory. In our algorithm we explore the case where the memory is not enough for in-memory aggregation, so disk spilling happens after the hash table is full. In our experiments we also observe significant hash miss cost in our Pre-Partitioning algorithm, and we use an optimized hash table design to solve this problem. 
\section{Processing Environment}\label{sec:proenv}

We now proceed to describe the main characteristics of the aggregation operation that we consider as well as the assumptions about the
data and resources used. 
\subsection{Aggregate Functions}
\label{sec:proenv:func}

Our focus is on aggregate functions \cite{bibd:techreport//epsteine1979/techniques-for-processing} such as aggregation combined with the ``GROUP-BY'' clause in SQL. As an example, 
consider the ``big data'' dataset \textbf{UserVisits} from \cite{DBLP:conf/sigmod/PavloPRADMS09}; it contains a visit history of web pages with the attributes shown in Table~\ref{tbl:uvattr}.

\begin{center}
\vskip -12pt
\begin{table}[ht]
\begin{tabular}{ll}
 Attribute Name                   &  Description                                \\
\hline
\verb|sourceIP| & the IP address (the source of the visit) \\
\verb|destURL| & the URL visited \\
\verb|adRevenue| & the revenue generated by the visit \\
\verb|userAgent| & the web client the user used \\
\verb|countryCode| & the country the visit is from \\
\verb|languageCode| & the language for the visit \\
\verb|searchWord| & the search keyword \\
\verb|duration| & the duration of the visit
\end{tabular}
\caption{Attributes in UserVisits dataset.}\label{tbl:uvattr}
\end{table}
\end{center}
\vskip -32pt
An example \verb|GROUP BY| aggregation appears in the following SQL query, which for each \verb|sourceIP| address (representing a unique user), computes the total advertisement revenue and the total number of visits:

\begin{Verbatim}[xleftmargin=5mm]
SELECT sourceIP, SUM(adRevenue), COUNT(*)
FROM UserVisits
GROUP BY sourceIP
\end{Verbatim}

The {\bf by-list} (the \verb|GROUP BY| clause in the example) specifies the \textbf{grouping key}, while the {\bf aggregate function(s)} (\verb|SUM| and \verb|COUNT| in the example) specify the way to compute the \textbf{grouping state}. The grouping state in the above example has two aggregated values (sum and count). The result of the aggregation (the \textbf{group record} or \textbf{group} for short) contains both the grouping key and the grouping state. 

Many commonly-used aggregate functions, like the \verb|SUM| and \verb|COUNT| in the example, can be processed in an accumulating way, i.e., for each group, only a single grouping state (with one or more aggregate values) needs to be maintained in memory, no matter how many records belong to this group. Similar \textbf{bounded-state} aggregate functions include \verb|AVERAGE|, \verb|MIN|, and \verb|MAX|. Many other aggregate functions, like finding the longest string of a given field per group (i.e., ``find the longest \verb|searchWord| for each \verb|sourceIP| in the \verb|UserVisits| dataset''), can be considered as bounded-state functions if the memory usage of the grouping state is bounded (for example the \verb|searchWord| could be at most 255 characters long, which is a common constraint in relational databases). A query with multiple aggregate functions on the same group-by condition is also bounded on the state, as far as each of them is a bounded-state function. So all our discussion in the paper also applies to this case.

However, there are aggregate functions that are not in the bounded-state category. An example is \verb|LISTIFY| (supported in AsterixDB) which for each group returns all records belonging to that group in the form of a (nested) list. 
Since the size of the grouping state depends on the number of group members, its memory usage could be unbounded. In this paper we concentrate primarily on bounded-state aggregate functions, as those are the most common in practice. Note that the simpler, scalar aggregation can be considered as an aggregate function with a single group (and thus all algorithms we will discuss can be applied to scalar aggregation directly).

\subsection{Data and Resource Characteristics}
\label{sec-4-2}

We assume that the size of the dataset can be far larger than the available memory capacity, so full in-memory aggregation could be infeasible. Whether our algorithms use in-memory or external processing depends on the total size of the grouping state, which is decided by the number of unique groups in the dataset (grouping key cardinality), and also the size of the grouping state.  An efficient aggregation algorithm should be able to apply in-memory processing if all unique groups can fit into memory, and shift dynamically to external processing otherwise. 




This paper assumes a commonly-used frame-based memory management strategy, which has been implemented in the Hyracks \cite{DBLP:conf/icde/BorkarCGOV11} data processing engine where all our algorithms are implemented. 
The Hyracks engine manages the overall system memory by assigning a tightly bounded memory budget to each query, in order to support parallel query processing. We will use $M$ to denote the memory budget (in frames or memory pages) for a particular aggregation query, $R$ to denote the size of the input data in frames, and $G$ the size of the result set in frames.


For aggregation algorithms that utilize a hash table, current Hyracks operators use a traditional separate chaining hash table with linked lists \cite{DBLP:books/aw/Knuth73}. The memory assigned to a hash table is used by a {\bf slot table} and a {\bf list storage area}. The slot table contains a fixed number of slots $H$ (i.e., it is static hashing; $H$ is also referred to as the {\bf slot table size}). Each non-empty slot stores a pointer to a linked list of group records (whose keys were hashed to that slot). The list storage area stores the actual group records in these linked list(s). Group records from different slots can be stored in the same frame. A new group is hashed into a slot by being inserted to the head of the linked list of that slot (or creating a new linked list if the slot was empty). An already-seen group is aggregated by updating its group record in the linked list. 

\begin{figure}[hbt]
\vskip -6pt
\centering
\epsfig{file=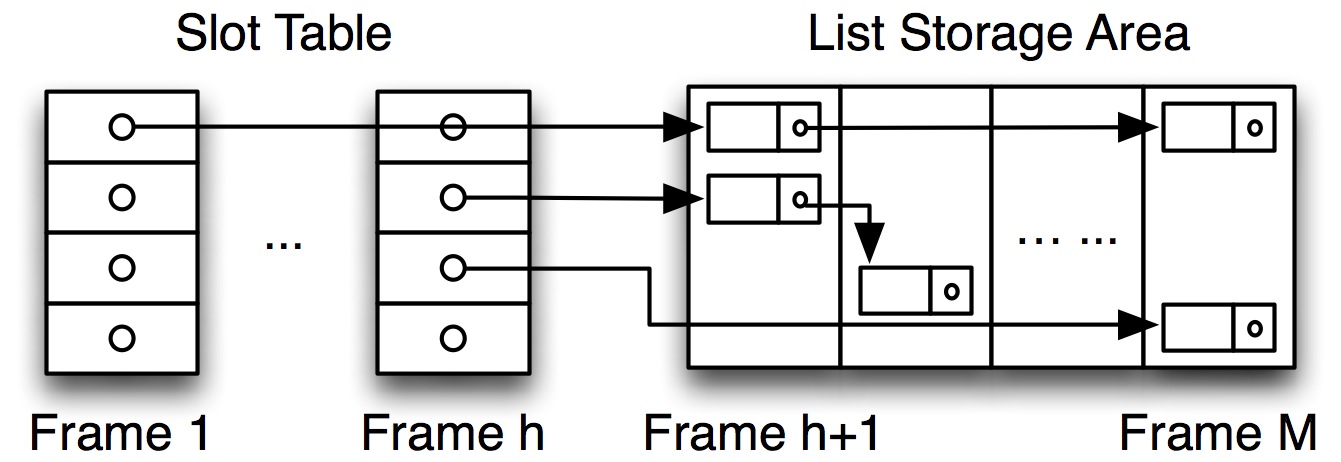, width=2.5in}
\caption{An In-memory Hash Table.}\label{fig:hashtable}
\vskip -12pt
\end{figure}
An in-memory hash table is {\bf full} when no new group record can fit in the list storage area based on its given memory budget. Figure~\ref{fig:hashtable} shows such an in-memory hash table with a budget of $M$ frames (for both the slot table and the list storage area), where $h$ frames are occupied by the slot table.

\section{Aggregation Algorithms}\label{sec:algs}

This section takes an in-depth look at six candidate aggregation algorithms: the Sort-based, the Hash-Sort, and four hybrid-hash-based algorithms (Original Hybrid-Hash, Shared Hashing, Dynamic Destaging, and Pre-Partitioning). The Hash-Sort and Pre-Partitioning algorithms are novel, while the others are based on adapting approaches discussed in the previous literature. Table~\ref{tbl:allalg} gives an overview of these algorithms. 

\subsection{Sort-based Algorithm}
\vspace{-3mm}
\begin{figure}[bt]
\centering
\epsfig{file=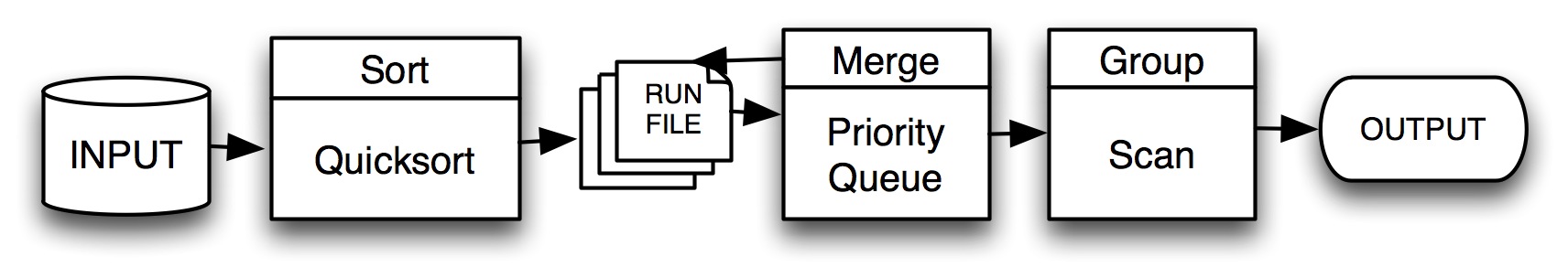, width=3.2in}
\caption{Sort-based Algorithm}\label{fig:sort-based}
\vskip -16pt
\end{figure}

The classic Sort-based aggregation algorithm includes two phases, \textbf{sort} and \textbf{aggregate}. Figure~\ref{fig:sort-based} depicts the algorithm's workflow. The sort phase sorts the data on the grouping key (using a sort-merge approach), while the aggregate phase scans the sorted data once to produce the aggregation result. In detail:

\begin{itemize}
\item Phase 1 (External Sort): (i) \textbf{Sort}: Data is fetched into memory in frames. When the memory is full, all in-memory records are sorted using the Quicksort algorithm \cite{DBLP:journals/siamcomp/Sedgewick77}, and flushed into a run file. If the total projected input data size is less than the memory size, the sorted records are maintained in memory and no run is generated. Otherwise, runs are created until all input records have been processed.
(ii) \textbf{Merge}: Sorted runs are scanned in memory, with each run having one frame as its loading buffer. Records are merged using a loser-tree \cite{DBLP:books/aw/Knuth73}. If the number of runs is larger than the number of available frames in memory, multiple levels of merging are needed (and new runs may be generated during the merging). 

\begin{table}[t]
\begin{tabular}{|c||c|c|}
\hline
Algorithm & Using Sort? & Using Hash? \\ \hline \hline
Sort-based \cite{bibd:techreport//epsteine1979/techniques-for-processing},\cite{DBLP:journals/tods/BittonBDW83},\cite{DBLP:books/daglib/0011128}, \cite{DBLP:books/daglib/0015084} & Yes & No\\ \hline
Hash-Sort (New) & Yes & Yes \\ \hline
Original Hybrid-Hash \cite{DBLP:journals/tods/Shapiro86} & No & Yes \\ \hline
Shared Hashing \cite{DBLP:conf/sigmod/ShatdalN95} & No & Yes \\ \hline
Dynamic Destaging \cite{DBLP:conf/vldb/GraefeBC98} & No & Yes \\ \hline
Pre-Partitioning (New) & No & Yes \\ \hline
\end{tabular}
\caption{Overview of all six algorithms.}\label{tbl:allalg}
\vskip -12pt
\end{table}

\item Phase 2 (Group): 
Each output record of the last round of merging in Phase 1 (i.e., when the number of runs is less than or equal to the available frames in memory) will be aggregated on-the-fly, by keeping just one group as the current running group in memory and comparing the merge output record with the running group: if they have the same grouping key,  they are aggregated; otherwise, the running group is flushed into the final output and replaced with the next merge output record. This continues until all records outputted from Phase 1 are processed.

\end{itemize}

\vspace{-3mm}
\begin{figure}[hbt]
\centering$\begin{array}{cc}
\epsfig{file=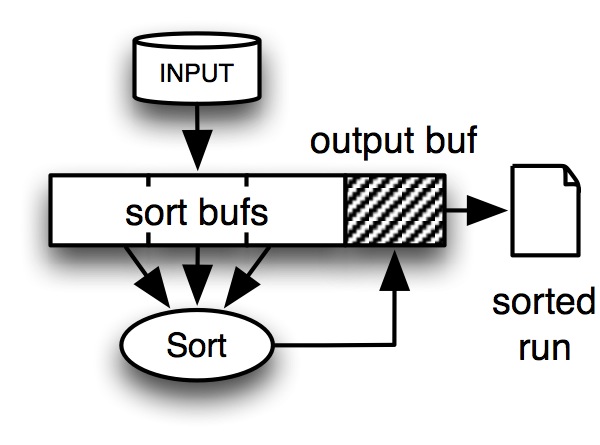, height=1.0in} &
\epsfig{file=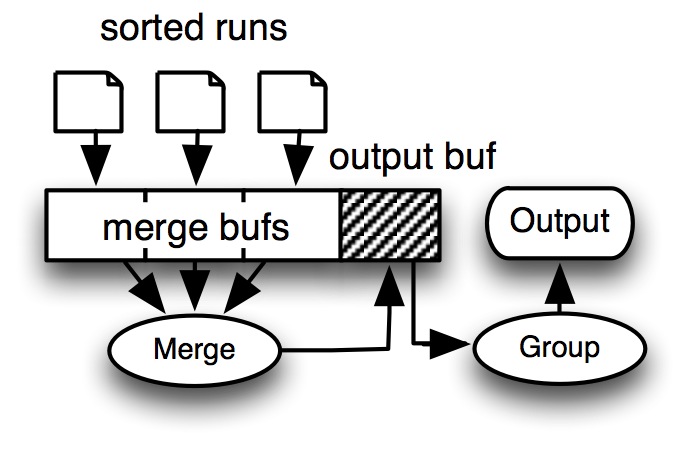, height=1.05in} \\
(a)\mbox{ Sort in Phase 1} & (b) \mbox{ Group in Phase 2}
\end{array}$
\caption{Memory structure in the Sort-based algorithm.}\label{fig:sort-based-mem}
\vskip -6pt
\end{figure}

The algorithm uses only the available memory budget $M$, since (i) the in-place Quicksort algorithm \cite{DBLP:journals/siamcomp/Sedgewick77} sorts $M - 1$ frames with one frame as the output buffer, and (ii) for merging, at most $M - 1$ runs will be merged in a single merge round, and multiple-level merging will ensure this if the number of runs is larger than $M - 1$. The group phase is pipelined with the last round of merging and it needs to maintain only one running group in memory (since the input records of this phase are provided in sorted order on the grouping key).


\subsection{Hash-Sort Algorithm}\label{sec:alg:hashsort}

The main disadvantage of the Sort-based algorithm is that it first scans and sorts the whole dataset. For a dataset that can be collapsed during aggregation, applying aggregation at an early stage would potentially save both I/O and CPU cost. The Hash-Sort algorithm that we developed for AsterixDB takes advantage of this observation by performing some aggregation before sorting. Figure~\ref{fig:hashsort-based} illustrates the workflow of this algorithm. Specifically, the Hash-Sort algorithm contains two phases, as described below:

\vspace{-3mm}
\begin{figure}[hbt]
\centering
\epsfig{file=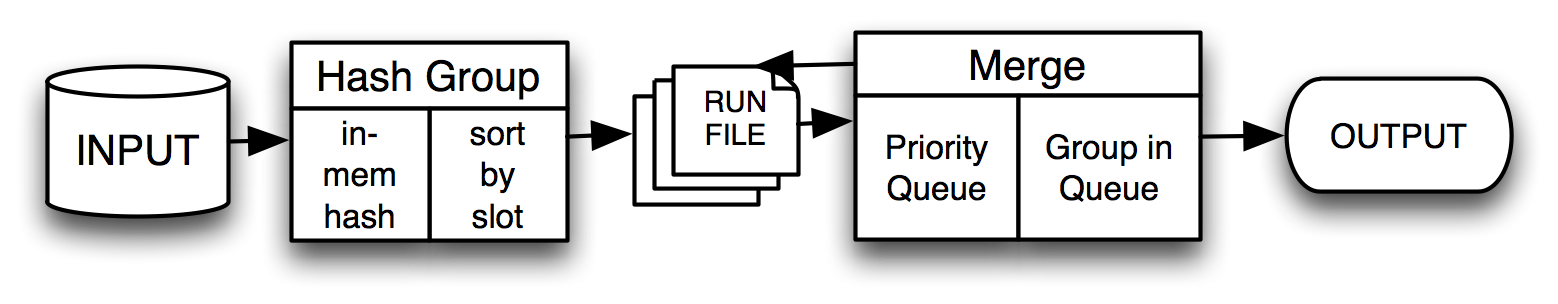, width=3in}
\caption{Hash-Sort Algorithm}\label{fig:hashsort-based}
\vskip -6pt
\end{figure}

\begin{itemize}
\item Phase 1 (Sorted Run Generation): An in-memory hash table is initialized using $M-1$ frames while the remaining frame is used as an output buffer. Input records are hashed into the hash table for aggregation. A new grouping key creates a new entry in the hash table, while a grouping key that finds a match is aggregated. 
When the hash table becomes full, the groups within each slot of the table are sorted (per slot) on the grouping key using in-place Quicksort, and the sorted slots are flushed into a run file in order of slot id (i.e., records in each run are stored in (\verb|slot-id|, \verb|grouping key|) order). The hash table is then emptied for more insertions. This continues until all input records have been processed. If all groups manage to fit into the hash table, the table is then directly flushed to the final output (i.e., Phase 2 is not applicable).
\item Phase 2 (Merge and Group): Each generated run is loaded using one frame as its loading buffer, and an in-memory loser-tree priority queue is built on the combination of (\verb|slot-id|, \verb|grouping key|) for merging and aggregation. The first group record popped is stored in main memory as the running group. If the next group popped has the same grouping key, it is aggregated. Otherwise, the running group is written to the output and is replaced by the new group (just popped).
This process continues until all runs have been consumed. Similar to the Sort-based algorithm, at most $M - 1$ runs can be merged in each round; if more runs exist, multiple-level merging is employed.
\end{itemize}

\vspace{-3mm}
\begin{figure}[hbt]
\centering$\begin{array}{cc}
\epsfig{file=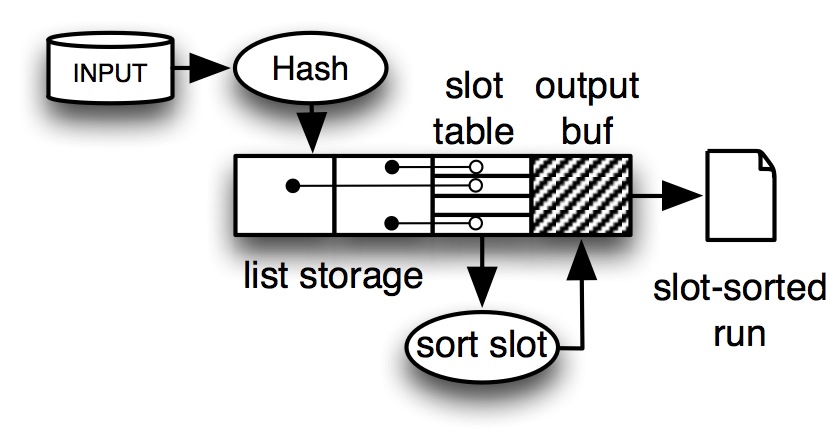, height=1.0in} &
\epsfig{file=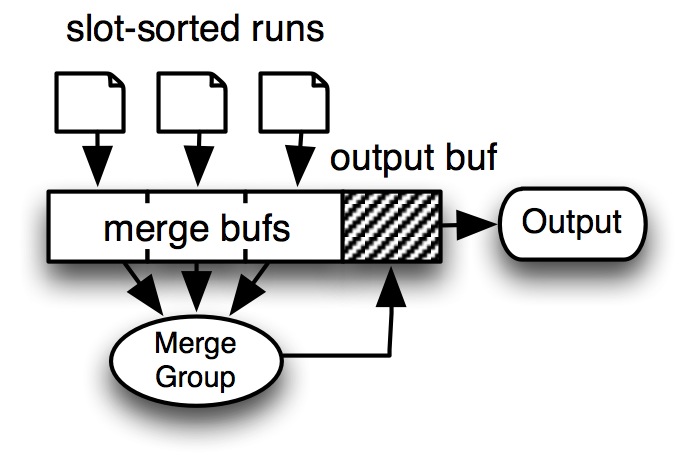, height=1.05in} \\
(a)\mbox{ Phase 1} & (b) \mbox{ Phase 2}
\end{array}$
\caption{Memory structure in the Hash-Sort algorithm.}\label{fig:hashsort-mem}
\vskip -6pt
\end{figure}

This algorithm also uses a bounded memory budget. Figure~\ref{fig:hashsort-mem} shows the memory configuration in its two phases. In the first phase the in-memory hash table uses exactly $M-1$ frames of the memory, and the table is flushed and emptied when it is full. Sorting (although slot-based) and merging are similar to the Sort-based algorithm in terms of memory consumption. 

\subsection{Hybrid-Hash Variants}
\label{sec:algs:hybridhash}

Hybrid-hash algorithms assume that the input data can eventually be partitioned so that one partition (the {\bf resident partition}) can be completely aggregated in-memory, while each of the other partitions ({\bf spilling partitions}) is flushed into a run and loaded back later for in-memory processing. I/O is thus saved by avoiding writing and re-reading the resident partition. 
Specifically, there are $(P+1)$ partitions created, with the resident partition (typically partition 0) being aggregated in-memory using $M-P$ frames, and the other $P$ partitions being spilled using $P$ frames as their output buffers. The required number of spilling partitions $P$ can be calculated for a given memory budget $M$ assuming that (i) the full memory can contain an in-memory hash table for the resident partition plus one frame for each spilling partition, and (ii) the size of each spilling partition is bounded by the memory size (and can thus be processed in-memory in the next step). The following formula gives a formal description of this partition strategy:

\vspace{-2mm}
\begin{align}\label{equ:hybridHashParts}
M - P &= G * F - (M - 1) * P \nonumber \\
\Rightarrow P &= {{G * F - M}\over{M - 2}} 
\end{align}

where $F$ is a fudge factor used to reflect the overhead from both the hash table and other structures (more about the fudge factor will be discussed in Section~\ref{sec:eval:hash}). This formula indicates that the total input data is processed as one resident partition (occupying $M - P$ frames for in-memory aggregation), and $P$ spilled partitions (each will fit into memory using $M-1$ frames). The above formula appeared in \cite{DBLP:journals/tods/Shapiro86} for joins; we adapt it here for aggregation, so it uses the result set size $G$ instead of the input size $R$, because records from the same group will be aggregated and collapsed.

\begin{figure}[hbt]
\centering
\psfig{file=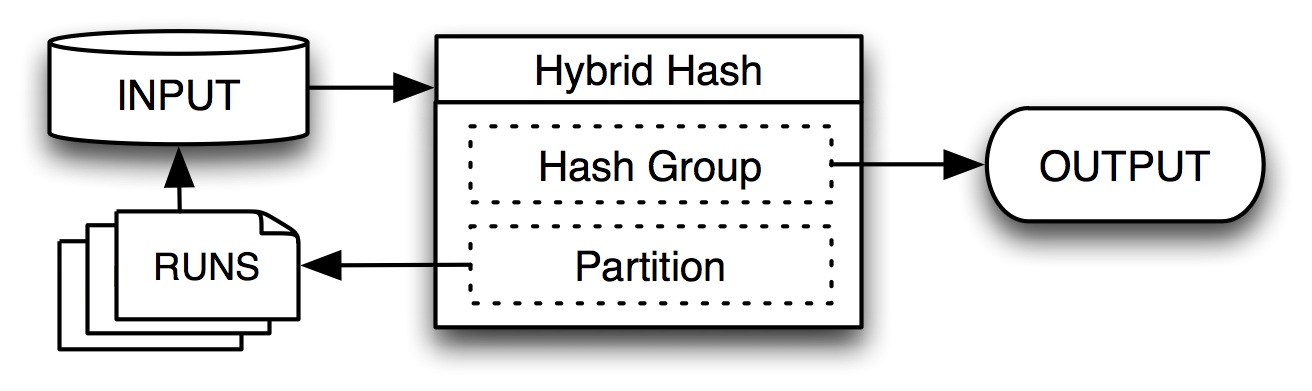, height=0.8in}
\caption{General Hybrid-hash algorithm structure.}\label{fig:hybridhash}
\vskip -6pt
\end{figure}

All hybrid-hash algorithms in this paper process data recursively using two main phases as illustrated in Figure~\ref{fig:hybridhash}. In detail, 




\vspace{-3mm}

\begin{itemize}
\item Phase 1 (Hash-Partition): If the input data is too large to be hybrid-hash processed in one pass ($G\geq M^2$), all memory is used to partition the input data into smaller partitions ({\bf grace partitioning} in Grace Join \cite{DBLP:journals/ngc/KitsuregawaTM83}), and for each partition one run file is generated. Otherwise ($G < M^2$), partition 0 is immediately aggregated using an in-memory hash table. At the end of this phase, partition 0 will be either flushed into a run file (if its aggregation is not completed due to the incorrect estimation on partitioning) or directly flushed to the final output (otherwise).
\item Phase 2 (Recursive Hybrid-Hash): Each run file generated above is recursively processed by applying the hash-partition algorithm of Phase 1. The algorithm terminates if there are no runs to be processed. To deal at runtime with grouping key value skew, if a single given run file's output is more than 80\% of the input file that it was partitioned from, or the number of  grace partitioning levels exceeds the number of the levels that would have been needed for the Sort-based algorithm, this particular run file will be processed next using the Hash-Sort algorithm (instead of recursive hybrid-hash) as a fallback to avoid deep recursion. 
\end{itemize}

Clearly, hybrid-hash algorithms need $G$ as an input parameter in order to manage memory space optimally. While the aim is to fully aggregate the resident partition in memory (when $G < M^2$), this is not guaranteed under strictly bounded memory by the existing hybrid-hash approaches we have seen, including the Original Hybrid-Hash \cite{DBLP:journals/tods/Shapiro86}, the Shared Hashing \cite{DBLP:conf/sigmod/ShatdalN95} and the Dynamic Destaging \cite{DBLP:conf/vldb/Graefe99} algorithms.
Thus in AsterixDB we propose a new approach using Pre-Partitioning that guarantees the completion of resident partition in memory. The details of these variants are described in the following subsections.

\subsubsection{Original Hybrid-Hash}

\begin{figure}[hbt]
\centering
\epsfig{file=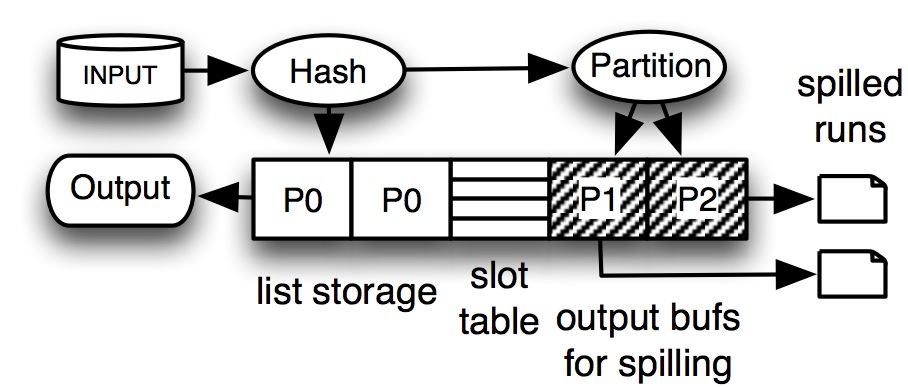, height=1.0in}
\caption{Memory Structure in the Original Hybrid-hash Algorithm.}\label{fig:originhybridhash}
\vskip -6pt
\end{figure}

In this algorithm, adapted from \cite{DBLP:journals/tods/Shapiro86}, if an input record is hashed to partition 0, then it is inserted into the in-memory hash table for aggregation, otherwise it is moved to an output buffer for spilling. Figure~\ref{fig:originhybridhash} depicts the memory structure of this algorithm. Ideally partition 0 should be completely aggregated in-memory and directly flushed to the final output; however if the hash table becomes full, groups in the list storage area are simply flushed into a run (i.e., partition 0 also becomes a spilling partition). 

\vspace{-3mm}

Note that the proper choice of the number of spilled partitions $P$ depends on the result size $G$ which is unknown and can only be estimated. An incorrect estimation of $G$ may result in partition 0 being too large to fit into memory and finally being spilled. While this may cause more I/O, the memory usage of this algorithm is still tightly bounded, since at most $M$ frames are used during the whole procedure. 


\subsubsection{Shared Hashing}
\label{sec:alg:hybridhash:sharedhashing}

The hybrid-hash algorithm proposed in \cite{DBLP:conf/sigmod/ShatdalN95} creates the same partitions as the Original Hybrid-Hash does, but the in-memory hash table is {\bf shared} by all partitions. This sharing allows for aggregating data from both partition 0 {\bf and} the other $P$ partitions. Effectively, the Shared Hashing algorithm initially treats all partitions as `resident' partitions. In order to use as much of memory for aggregation for all partitions, and also to reserve enough output buffers for spilling partitions, the list storage area of the hash table is divided into two parts: the {\bf non-shared} part contains $P$ frames for the $P$ spilling partitions, while the remaining frames  ({\bf shared} part) are assigned to partition 0 but initially shared by all partitions. Using this layout, the $P$ frames for spilling partitions can also be used for hashing and grouping before the memory is full, and then for spilling output buffers after that. Figure~\ref{fig:sharedhash1} illustrates the memory structure of this stage. 
\verb|P1| and \verb|P2| are the frames allocated to partition 1 and 2 respectively so they are not shared. Other frames (marked as \verb|Px|) are assigned to partition 0, but also shared by partition 1 and 2 before any spilling.
Spilling is triggered when a new group record arrives to one partition and there is no space available for more data from that partition (including the shared frames). The first two spillings are handled differently from future ones, as described below:

\vspace{-3mm}
\begin{figure}[hbt]
\centering
\psfig{file=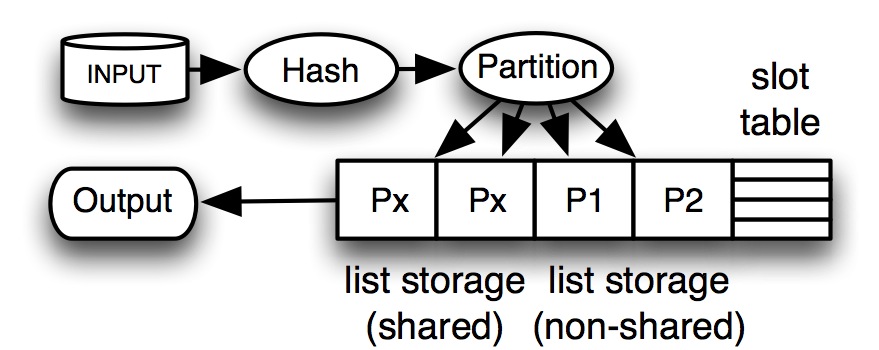, height=1in}
\caption{Memory structure before the first spilling of the Shared Hashing algorithm.}\label{fig:sharedhash1}
\vskip -6pt
\end{figure}

\begin{itemize}

\item First Spilling: When the first spilling is triggered (from any partition) by lack of additional space, all $P$ spilling partitions are flushed. 
Each frame in the (soon-to-be) non-shared part is first flushed into a run for its corresponding partition using partition 0's output buffer. After flushing, the non-shared frames will become the output buffers for the $P$ spilling partitions. Then the shared part is scanned. Group records from all spilling partitions are moved to the corresponding partition's output buffer for spilling, while groups of partition 0 are rehashed into a new list storage area built upon recycled frames (i.e., a frame in the shared part is recycled when all its records have been completely scanned and moved) and clustered together. Figure~\ref{fig:sharedhash2} depicts the memory structure when the scan is processed. After the first spilling the new list storage area belongs only to partition 0, and there is one output buffer for each spilling partition; the memory structure is now the same as the Original Hybrid-Hash showed in Figure~\ref{fig:originhybridhash}.

\vspace{-3mm}
\begin{figure}[hbt]
\centering
\psfig{file=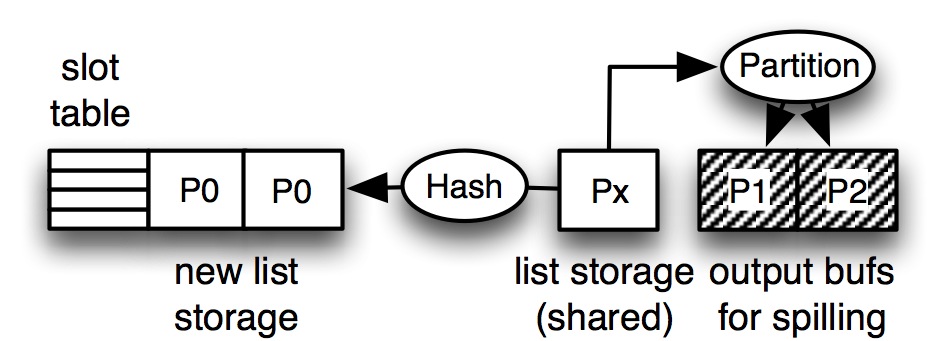, height=0.9in}
\caption{Memory structure during the first spilling of the Shared Hashing algorithm.}\label{fig:sharedhash2}
\vskip -6pt
\end{figure}


\item Second Spilling: When the new list storage area has no more space for new group records from partition 0, partition 0 will be spilled. Its groups are flushed to a run file, and the frames they occupied are recycled. From now on, a single frame is reserved as the output buffer for partition 0 as well, and it is directly spilled like the other partitions.

\end{itemize}

The above algorithm uses bounded memory. Before the first spilling, the entire $M$-frame memory allocation is used as an in-memory hash table. When scanning the shared part in the first spilling, non-shared frames are reserved for the $P$ spilling partitions, and frames recycled from the shared part are used for the new list storage area to cluster the partition 0 groups. After the second spilling the out-buffer frames for spilling partitions are obviously always memory-bounded.

\subsubsection{Dynamic Destaging}
\label{sec:alg:hybridhash:dynamicdestaging}

\begin{figure}[hbt]
\centering$\begin{array}{c}
\epsfig{file=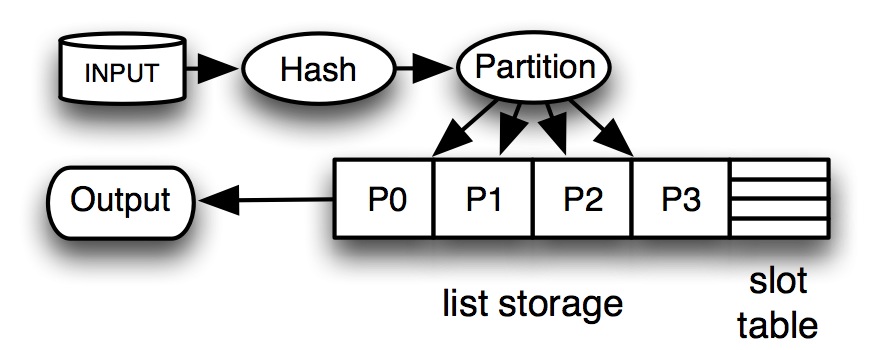, height=0.9in}\\
(a) \mbox{ Before spilling (all partitions are resident).}\\
\epsfig{file=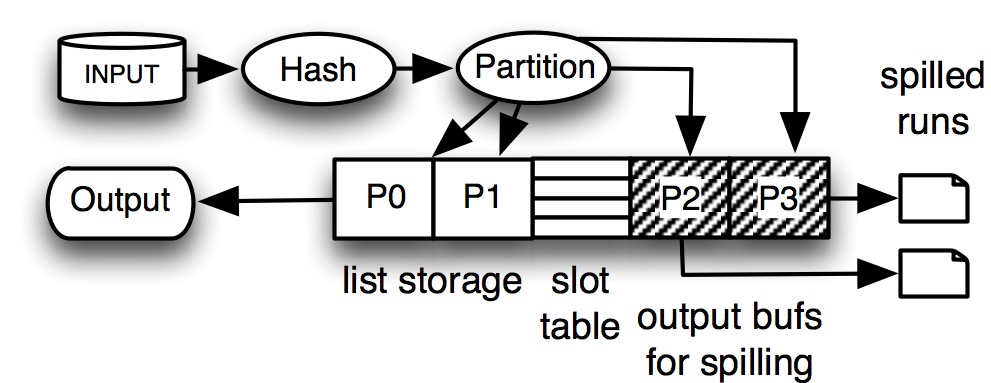, height=0.9in}\\
(b) \mbox{ After partition 2 and 3 are spilled.}
\end{array}$
\caption{Memory structure in the Dynamic Destaging algorithm.}\label{fig:dynamicdestaging}
\vskip -6pt
\end{figure}

Unlike the previous two approaches, where the memory space for partition 0 is pre-defined (based on Formula~\ref{equ:hybridHashParts}), the Dynamic Destaging algorithm \cite{DBLP:conf/vldb/Graefe99} dynamically allocates memory among all partitions, and spills the largest resident partition when the memory is full. After all records have been processed, partitions that remain in memory can be directly flushed to the final output (i.e., they are all resident partitions). This algorithm has two stages:

\begin{itemize}
\item Stage 1 (Initialization): An in-memory hash table is built so that one frame is reserved for the resident partition and each of the $P$ spilled partitions in the list storage area, and the remaining frames are managed in a buffer pool. All partitions are initially considered to be resident partitions. Figure~\ref{fig:dynamicdestaging} (a) depicts the memory structure after this stage.
\item Stage 2 (Hash-and-Partition): Each input record is hashed and aggregated into the frame of the corresponding partition. When a frame becomes full, a new frame is allocated from the pool for this partition to continue its aggregation. If no frame can be allocated, the largest (still) resident partition is spilled into a run file. Frames that this partition occupied are recycled, and a single frame is now reserved as its output buffer. Additional records hashed to such a spilled partition will be directly copied to its output buffer for spilling (i.e., no aggregation happens for a spilled partition and no additional frames will be allocated for that partition in the future). Figure~\ref{fig:dynamicdestaging} (b) illustrates the memory structure after some partitions are spilled.
\end{itemize}

\vspace{-3mm}

Following \cite{DBLP:conf/vldb/Graefe99}, for computing the initial number of spilled partitions $P$, our implementation allocates between 50-80\% of the available memory (i.e., 50\% if the computed $P$ value is less than 50\% and 80\% if the computed $P$ is larger than 80\%), in order to balance the size of the in-memory and spilling partitions (i.e., so the partition size is not too large or too small). Small runs created due to possible over-partitioning are merged and processed together in a single in-memory hash aggregation round, if the merged run can be fully processed in-memory (mentioned as {\bf partition tuning} in \cite{DBLP:conf/vldb/Graefe99}).

The Dynamic Destaging algorithm is memory bounded since memory is dynamically allocated among all partitions. When memory becomes full, a partition is spilled to recycle space. In the worst case, when all partitions are spilled, the available memory can be dynamically allocated among all partitions and used simply as output buffers.

\subsubsection{Pre-Partitioning}
\label{sec:algs:hybridhash:prepartitioning}

All of the approaches described so far assume that the hash function and the distribution of the hash values into partitions are properly chosen so that resident partition(s) can be completely aggregated in memory. Unfortunately, there can be no such guarantee, especially without precise knowledge about the input data. The naive approach of partitioning the hash value space based on Formula~\ref{equ:hybridHashParts} will not work if the hash values used by the input data are not uniformly distributed in the hash value space.
Moreover, these hybrid-hash aggregation algorithms are all derived from (and thus influenced by) hybrid-hash joins. One important property that distinguishes aggregation from join is that, in aggregation, the size of a group result is fixed and is not affected by duplicates. As a result, the memory requirement for a set of groups is fixed by the cardinality of the set (while the group size in a join could be arbitrarily large). 

Based on these observations, we developed and implemented in AsterixDB the Pre-Partitioning algorithm. This algorithm divides the entire memory space similarly to the Original Hybrid-Hash, where $M-P$ frames are used for an in-memory hash table for partition 0. But, instead of assigning the groups of partition 0 based on hash-partitioning, Pre-partitioning considers all groups that can be inserted into the in-memory hash table (before the table becomes full) as belonging to partition 0. 

\begin{figure}[hbt]
\centering
\psfig{file=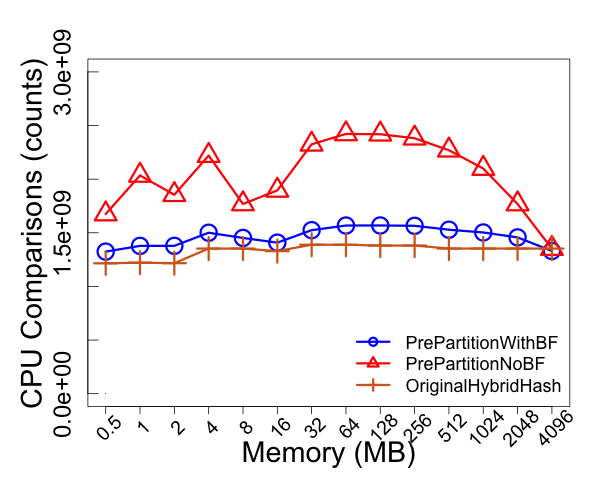, height=1.6in}
\caption{Comparisons of CPU cost among Pre-Partition with bloom filter, Pre-Partition without bloom filter, and the Original Hybrid-Hash.}\label{fig:prepartbf}
\vskip -6pt
\end{figure}

After the hash table is full, grouping keys that cannot be aggregated in the in-memory partition are spilled into the remaining $P$ output frames. 
In order to decide whether a record should be spilled or aggregated, each input record needs a hash table lookup to check whether it can be aggregated or not. This would cause a much higher hash lookup miss ratio compared with other hybrid-hash algorithms. To improve the efficiency of identifying the memory-resident v.s. spilling groups, we add an extra byte as a mini bloom filter for each hash table slot. The bloom filter is updated when a new group is inserted into the slot (before the hash table becomes full). After the hash table is full, for each input record a lookup on the bloom filter is first performed, making a hash table lookup necessary only when the bloom filter lookup returns true. If the bloom-filter lookup returns false, it is safe to avoid looking into the hash table (since a bloom filter could only cause a false-positive error). For a properly sized hash table (i.e. where the number of slots is no less than the number of groups that can be contained in the table), the number of groups in each slot will be small (less than two on average), and a 1-byte bloom-filter per slot works well to reduce hash table lookups with a very low false-positive error rate. Figure~\ref{fig:prepartbf} shows the CPU cost of aggregating 1 billion records with around 6 million unique groups using Pre-Partitioning with bloom filtering, Pre-Partitioning without bloom filtering, and the Original Hybrid-Hash algorithms. From the figure we can see that by applying the bloom filter, the CPU cost of the Pre-Partitioning algorithm is greatly reduced and becomes very close to the cost of the Original Hybrid-Hash algorithm. 

In order to reduce the overhead of maintaining the bloom filters, in our implementation no bloom filter lookup is performed before the hash table is full. This means that there is only the cost of updating the bloom filters when updating the hash table through a negligible bit-wise operation. This is because before the hash table is full, all records are inserted into the hash table anyway, and the benefit from bloom filters on reducing the hash misses is very limited (since a hash miss because of an empty slot can be easily detected without bloom filter lookup). Furthermore, if the dataset could be aggregated in memory based on the input parameters, no bloom filter will be needed, and the bloom filter overhead can be eliminated. Note that the output key cardinality ($G$ in Formula~\ref{equ:hybridHashParts}) could be underestimated, and the bloom filters could be falsely disabled, causing more CPU cost on hash misses. Pre-Partitioning still outperforms other hybrid-hash algorithms in this case because other hybrid-hash algorithms have more extra I/O cost on spilling the in-memory partition. Section~\ref{sec:eval:hhinput} shows our experiments in this scenario.
Figure~\ref{fig:prepartitioning} shows the two stages of the Pre-Partitioning algorithm:

\vspace{-3mm}
\begin{figure}[hbt]
\centering$\begin{array}{c}
\epsfig{file=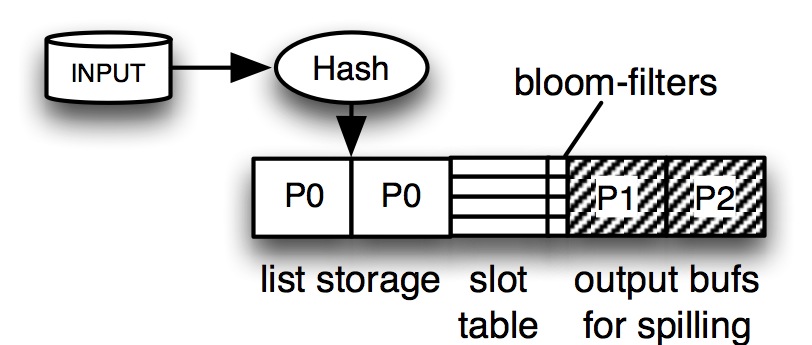, height=0.9in}\\
(a) \mbox{ Partition-0-build: before the hash table is full.}\\
\\
\epsfig{file=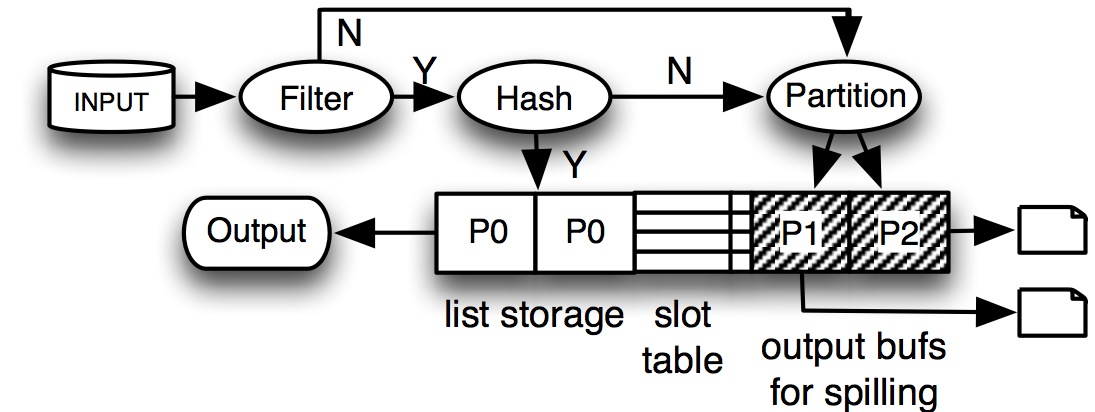, height=1.0in}\\
(b) \mbox{ Hash-And-Partition: after the hash table is full.}
\end{array}$
\caption{Memory Structure in the Dynamic Destaging Algorithm.}\label{fig:prepartitioning}
\vskip -6pt
\end{figure}

\begin{itemize}
\item Stage 1 (Partition-0-Build): Using Formula~\ref{equ:hybridHashParts}, $P$ frames are reserved as the output buffers (to be used in Stage 2 for the spilling partitions). The remaining $M-P$ frames are used as an in-memory hash table storing groups of partition 0. Input records are inserted into this hash table for aggregation until the list storage area is full. If $P > 1$, a 1-byte bloom filter is used for each hash table slot, and all insertions to the hash table update the respective bloom filters. Figure~\ref{fig:prepartitioning} (a) shows the memory structure of this stage.
\item Stage 2 (Hash-And-Partition): After the hash table is full, for each input record we check if that record has been seen before in partition 0 by first performing a bloom filter lookup; if the bloom filter lookup is positive, a hash table lookup follows, and it is aggregated if a match is found (no more memory is needed for this aggregation). Otherwise, this record is stored into one of the $P$ output frames. When such a frame becomes full it is spilled. Figure~\ref{fig:prepartitioning} (b) illustrates this procedure. When all records have been processed, the groups aggregated in the in-memory hash table are directly flushed to the final output.
\end{itemize}

The Pre-Partitioning algorithm uses bounded memory since the in-memory hash table never expands beyond the $M-P$ pre-allocated frames. A benefit of this algorithm is that it allocates as many records to partition 0 as possible (until the in-memory hash table becomes full, at which time the pre-allocated $M-P$ frames are fully utilized) and this partition is guaranteed to be fully aggregated in-memory. Since the previous hybrid-hash variants cannot provide this guarantee, they may not fully utilize the pre-allocated memory for partition 0 (even if partition 0 could be finished in-memory). 

We have also explored the idea of applying the bloom filter optimization to other hash-based algorithms discussed in this paper. However the overhead would be more significant than the benefit for the other algorithms. This is because a bloom filter is useful to avoid hash collisions (i.e. using a bloom-filter may avoid the hash lookup leading to a hash miss). However, with properly sized hash tables and assuming good hashing functions, most of the hash table insertions will not cause a hash collision, so the bloom filter does not help much reduce the collisions but introduces more memory overhead for the hash table. 
\section{Cost Models}\label{sec:theo}

\begin{center}
\vskip -12pt
\begin{table}[t]
\begin{tabular}{ll}
 Symbol                   &  Description                                \\
\hline
 $b$                      &  Tuple size in bytes                        \\
 $o$			 & Hash table space overhead factor (for its slot \\
& table  and references of linked list) \\
  $p$                      &  Frame size in bytes                        \\
 $A$                      &  Collection of sorted run files generated                      \\
$\mathcal{D}(n, m)$ & Dataset with $n$ records and $m$ unique keys \\
 $G$                      &  Output dataset size in frames             \\
 $G_t$  &  Number of tuples in output dataset \\
  $H$                      &  Number of slots in hash table                    \\
 $K$			 & Hash table capacity in number of unique groups \\
 $M$                      &  Memory capacity in frames                      \\
 $R$                      &  Input dataset size in frames              \\
 $R_t$  &  Number of tuples in input dataset         \\
 $R_{H}$ &  Number of raw records inserted into a hash\\
& table before it becomes full \\
\end{tabular}
\caption{Symbols Used in Models}\label{tbl:symbols}
\end{table}
\end{center}

We proceed by introducing applicable cost models for all six aggregation algorithms discussed in this paper. For simplicity we assume that the grouping keys are uniformly distributed over the input dataset. Moreover, for the hybrid-hash algorithm models it is assumed that the input parameters (size of input file, number of unique keys etc.) are precise.
The analysis focuses on the CPU comparison cost (for sorting and hashing) and the I/O cost (read and write I/Os). 
For simplicity, we omit the CPU and I/O costs for scanning the original input file and flushing the final result since they are the same for all algorithms (any may be pipelined). We also omit the pointer swapping cost in sorting and merging since it is bounded by the comparison cost (for a random dataset, the swap count is around $1/3$ of the total comparisons \cite{DBLP:journals/siamcomp/Sedgewick77}).  

In our analysis, we use the following basic {\em component} models that are common for all algorithms, namely: the {\em input}, {\em sort}, {\em merge} and {\em hash} components. The details of these component models can be found in the Appendix. Table~\ref{tbl:symbols} lists the symbols used in the component and algorithmic models.

\begin{itemize}
    \item{{\bf Input Component}}: Let $\mathcal{D}(n, m)$ denote a dataset with total number of records $n$ and containing $m$ unique grouping keys. The model's input component computes the following two quantities:
    \begin{itemize}
        \item $I_{key}(r, n, m)$, denotes the number of unique grouping keys contained in $r$ records randomly drawn from $\mathcal{D}(n, m)$ without replacement (see Equation \ref{equ:key_dep}).
        \item $I_{raw}(k, n, m)$, denotes the number of random picks needed from $\mathcal{D}(n, m)$ in order to get $k$ unique grouping keys (see Equation \ref{equ:raw_dep}).
    \end{itemize}
    \item{{\bf Sort Component}}: $C_{sort}(n, m)$ represents the number of CPU comparisons needed (using quicksort) in order to sort $n$ records containing $m$ unique keys (see Equation \ref{equ:sort}).
    \item{{\bf Merge Component}}: $C_{CPU.merge}(A, M)$ and \newline $C_{IO.merge}(A, M)$ represent the CPU and I/O cost respectively, for merging a set of files $A$ using $M$ memory frames (see Section \ref{theo:comp:merge}).
    \item{{\bf Hash Component}}: Assume a hash table whose slot table has $H$ slots and whose list storage area can store up to $K$ unique keys (i.e. at most $K$ unique groups can be maintained in the list storage area). The hash component computes the following quantities:
    \begin{itemize}
        \item $H_{slot}(i, H, n, m)$ represents the number of occupied slots in a hash table with $H$ slots after inserting $i$ random records ($i \leq K$) taken from $\mathcal{D}(n, m)$ (see Equation \ref{equ:uslots}).
        \item $C_{hash}(n, m, K, H)$ represents the total comparison cost until filling up the list storage area of a hash table with $H$ slots and capacity $K$ if the records are randomly picked from $\mathcal{D}(n, m)$ (see Equation \ref{equ:hash}). Note that the hash table could become full before all records from $\mathcal{D}(n, m)$ are loaded.
        \item $C_{hash}(n, m, K, H, u)$ is again the total comparison cost for filling up the list storage area as above, but assumes that $\mathcal{D}(n, m)$ has been partially aggregated, and the partially aggregated part ($u$ unique records) are first inserted into the hash table before the random insertion.
    \end{itemize}
\end{itemize}

\subsection{Sort-based Algorithm Cost}
\label{sec:theo:sortbased}

The I/O cost for the Sort-based algorithm is solely due to external sorting since grouping requires just a single scan that is pipelined with merging. Let $R$ denote the number of frames in the input dataset.  The sort phase scans the whole dataset once using $R$ write I/Os to produce $A$ sorted runs (where $|A|=\frac{R}{M}$), each of size $M$, that are then merged. The total I/O cost is thus: $C_{IO} = R + C_{IO.merge}(A, M)$.

The CPU comparison cost $C_{comp}$ consists of the sorting cost before flushing the full memory into a run and the merging cost for merging all sorted runs. Hence:

\vspace{-2mm}
\begin{align}\label{equ:sortcpu}
C_{comp} =& |A| * C_{sort}(R_{mem}, I_{key}(R_{mem}, R_{t}, G_{t})) \nonumber \\
 &+ C_{CPU.merge}(A, M)
\end{align}

where $R_{mem}$ denotes the number of records that can fit in memory ($R_{mem} = \frac{Mp}{b}$, where $p$ is the frame size and $b$ is the input record size), $R_{t}$ is the number of records in the input dataset, and $G_{t}$ is the number of unique keys in the input dataset (which is the same as the number of tuples in the output dataset).
 

\subsection{Hash-Sort Algorithm Cost}
\label{sec-6-5-2}

The Hash-Sort algorithm applies early aggregation using hashing and slot-based sorting. In the first phase (Sorted Run Generation), the I/O cost arises from flushing the unique keys in the hash table
whenever it becomes full. Since the hash table uses the whole available memory $M$, its capacity is $K = {{Mp}\over{ob}}$ (note that $o$ is used to represent the memory overhead per record due to the hash table structure). The number of raw records inserted into the hash table until it becomes full is then: $|R_{H}| = I_{raw}(K, R_{t}, G_{t})$. 
Once the hash table is full, all unique keys would be flushed after being sorted by (slot id, hash id). There are totally ${{R_{t}}\over{|R_{H}|}}$ files generated, each file with size ${{Kb}\over{p}}$; hence: $C_{IO.phase1} = {{R_{t}}\over{|R_{H}|}} * {{Kb}\over{p}}$.

The comparison cost for the first phase contains both hashing and slot-based sorting comparisons. The hashing comparison cost can be computed as 

\vspace{-2mm}
\begin{align}
C_{comp.hash} =& \frac{R_{t}}{|R_{H}|} * C_{hash}(|R_{H}|, G_{t}, K, H)
\end{align}

To estimate the sorting comparisons we note that when $K$ unique keys have been inserted, the number of
non-empty slots used is given by $H_{u}(|R_{H}|, \allowbreak H, R_{t}, G_{t})$. Based on the uniform distribution assumption, the
number of unique keys in each slot is: 
${L_{slot} = {{K}\over{H_{u}(|R_{H}|, H, R_{t}, G_{t})}}}$. Since duplicates have been aggregated, the $L_{slot}$ records to be sorted in each slot are all unique; hence the total number of comparisons due to sorting becomes:

\vspace{-2mm}
\begin{align}
C_{comp.sort} =& H_{u}(|R_{H}|, H, R_{t}, G_{t}) * C_{sort}(L_{slot}, L_{slot})
\end{align}

During the merging phase, the I/O cost includes the I/O for loading the sorted runs, and the I/O for flushing the merged file. The size of each sorted run generated by the sorting phase is the memory size $M$. The size of a merged file can be computed as the size (number) of unique keys contained in the sorted runs that are used to generate this merged file. The number of unique keys can be computed using the input component, given the number of raw records that are aggregated into the merged file. If $A$ denotes the total sorted runs and $A'$ denotes the files to be merged ($A' \subseteq A$ and $|A'| \leq M$), the number of raw records that will be aggregated into the merged file will be ${{R_{t} * |A'|}\over{|A|}}$, so the number of unique keys in the merged file would be $I_{key}({{R_{t} * |A'|}\over{|A|}}, R_{t}, G_{t})$. So the total I/O cost for merging the $AÕ$ sorted runs is

\vspace{-2mm}
\begin{align}
F(A') =& |A'| * M + {{I_{key}({{R_{t} * |A'|}\over{|A|}}, R_{t}, G_{t}) * b}\over{p}}
\end{align}

By applying $F(A')$ in $C_{IO.merge}(A, M)$, we can compute the total I/O cost for merging. The CPU comparison cost of merging the $A$ run files $C_{comp.merge}(A, M)$ can be computed in a similar way using the merge component. 


\subsection{Hybrid-Hash Based Algorithm Costs}
\label{sec:theo:hybridhash}

In this section we describe the cost model for the hash-partition phase (Phase 1) for each of the four hybrid-hash algorithms described in Section~\ref{sec:algs}. In the recursive hybrid-hash phase (Phase 2), all algorithms recursively process the produced runs using their hash-partition algorithm, and their cost can be easily computed by simply applying the cost model from Phase 1 so we omit the details.
When the key cardinality of the input dataset is too large for direct application of a hybrid-hash algorithm we need first to perform a simple partitioning until the produced partitions can be processed using hybrid hash. The cost of this partitioning is $2*L*R$ for its I/O cost of loading and flushing, and $L*R_{t}$ for CPU cost of scanning, if $L$ levels of partitioning are needed. 


\subsubsection{Original Hybrid-Hash}
\label{sec:theo:hybridhash:originhybridhash}

The Original Hybrid-Hash algorithm aggregates records from partition 0 only in its hash-partition phase while the other $P$ partitions are directly spilled using $P$ output buffers. Hence the available memory for the hash table is $(M - P)$ and the capacity of the hash table is $K = \frac{(M-P)p}{ob}$. Assuming that keys are uniformly distributed in the input dataset, partition 0 can be fully aggregated in the hash table. Since the number of raw input records of partition 0 is $\frac{K}{G_{t}} * R_{t}$, the comparison cost for hashing is $C_{hash}(\frac{K}{G_{t}} * R_{t}, K, K, H)$ (since the $\frac{K}{G_{t}} * R_{t}$ records contain $K$ unique keys). The I/O cost arises from loading the input records from the disk, and from spilling the raw records belonging to the $P$ spilled partitions onto the disk; hence $C_{IO} = R + (R - \frac{K}{G_{t}} * R)$. 




\subsubsection{Shared Hashing}
\label{sec:theo:hybridhash:static}%
The uniform key distribution and precise input parameter assumptions made by our cost model eliminate the second spilling phase of the Shared Hashing algorithm; hence the following discussion concentrates on the first spilling phase.
The Shared Hashing algorithm aggregates records from all partitions until the hash table is full. At this stage all memory except for one output buffer frame is used for the hash table, so the hash table capacity is $K = \frac{(M - 1)p}{ob}$. The hash comparison cost is thus similar to the Hash-Sort algorithm, i.e., $C_{comp.before\_full} = C_{hash}(|R_{H}|, G_{t},\allowbreak K, H)$. 


During the first spilling, grouping keys of partition 0 that are already in the hash table are re-hashed in order to be clustered together in a continuous memory space. Remaining records of partition 0 are hashed and aggregated until the hash table becomes full again. The fraction of partition 0 (the resident partition) $r_{res}$ and a spilled partition $r_{spill}$ in the total input dataset can be computed based on Formula~\ref{equ:hybridHashParts} as below:
\vspace{-2mm}
\begin{align*}
r_{res} = &\frac{M-P}{(M-P)+MP}, ~~r_{spill} = \frac{1 - r_{res}}{P}
\end{align*}
The hash comparison cost after the first spilling (including re-hashing and inserting the remaining records from partition 0) can be computed by considering that the ${r_{res} * K}$ unique groups are inserted ahead:

\vspace{-2mm}
\begin{align}\label{equ:comp_after_full}
C_{comp.after\_full} =& C_{hash}(r_{res}*R_{t} - I_{raw}(r_{res}*K, R_{t}, G_{t}), \nonumber \\
& r_{res} * G_{t}, K, H, r_{res}*K)
\end{align}

where $I_{raw}(r_{res}*K, R_{t}, G_{t})$ is the number of raw records inserted before the first spilling, while $r_{res} * K$ corresponds to the unique keys inserted before the first spilling that are then re-hashed during the first spilling. Here all partition 0 records are drawn from the $r_{res} * G$ unique keys assigned to partition 0.



After partition 0 is completely aggregated in memory and when the spilled runs are recursively processed, each run may already be partially aggregated, which corresponds to the `mixed' input case. Hence the comparison cost for the all resident partitions phase is computed as:
\vspace{-2mm}
\begin{align}
C_{comp.spill\_parts} =& C_{hash}(r_{spill} * R_{t} - \nonumber \\
&I_{raw}(r_{spill} * K, r_{spill} * R_{t}, r_{spill} * G_{t}),\nonumber \\
&r_{spill} * G_{t}, K', H, r_{spill} * K)
\end{align} 

This is very similar to the cost model showed in Equation \ref{equ:comp_after_full}, where the records inserted before the first spilling $(I_{raw}(r_{spill} * K, r_{spill} * R_{t}, r_{spill} * G_{t}))$ are collapsed into $(r_{spill} * K)$ unique records and reloaded during the recursive hashing.

The I/O cost emanates from the spilling partitions only. Since part of each spilling partition has been aggregated before the table is full, the I/O cost contains the I/O both for spilling the partially aggregated partition, and for flushing the remaining raw records of that partition (computed by subtracting the aggregated raw records from the total raw records of the spilling partition): 
\vspace{-2mm}
\begin{align}
C_{IO.spill} =& \frac{r_{spill}*K * b}{p} + r_{spill}*R\\
& - \frac{I_{raw}(r_{spill}*K, R_{t}, G_{t})*b}{p} \nonumber
\end{align}

where $\frac{r_{spill}*K * b}{p}$ is the I/O for spilling the partial aggregated results, and the remaining part is the I/O for spilling the raw records (where the records that are partially aggregated are excluded).





\subsubsection{Dynamic Destaging}
\label{sec:theo:hybridhash:dynamicdestaging}%
Until the hash table becomes full, the Dynamic Destaging algorithm behaves similarly to the Hash-Sort algorithm; hence the CPU comparison cost before any partition is spilled can be computed by $C_{hash}(|R_{H}|, G_{t}, K, H)$ (note that when this model is recursively applied to runs that have partially aggregated records, the `mixed' input Equation \ref{equ:hash_preinsert} should be used). 
When the hash table is full, the largest resident partition is spilled. The uniform assumption of the input dataset implies that at this time all partitions have the same number of grouping keys in memory; hence, any one partition can be randomly picked for spilling. If partition $i$ is picked for the $i$-th spill, the total available memory for the hash table is $M - (i - 1)$ (where $i - 1$ frames are used as the output buffers for the spilled partitions). The number of in-memory aggregated groups of the $i$-th spilling partition can be computed using Formula~\ref{equ:hybridHashParts} as:
\vspace{-2mm}
\begin{align*}
K_{i} =& {{K * (M - (i - 1))}\over{M(P + 1 - (i - 1))}}
\end{align*}
while the size of raw records hashed into the hash table for the $i$-th spilled partition is given by:
\vspace{-2mm}
\begin{align*}
R_{H.i} =& I_{raw}(K_{i}, {{R_{t}}\over{P+1}}, {{G_{t}}\over{P+1}})
\end{align*}

Note here that for a specific partition $i$, the hash table capacity and the number of slots are the portion of the total $K$ and $H$ assigned to this partition. Then the CPU comparison cost for hashing this partition becomes:
\vspace{-2mm}
\begin{align}\label{eqn:dd_comp_i}
C_{comp.i} =& C_{hash}(R_{H.i}, K_{i}, \frac{K}{P+1}, \frac{H}{P+1})
\end{align}

When spilling the $i$-th partition, since part of the partition has been hashed and collapsed before the partition is spilled, the total spilling I/O emanates from the raw records directly flushed $({{R}\over{P+1}} - R_{H.i})$, plus the partially aggregated unique keys $({{K_{i} * b}\over{p}})$; hence:

\vspace{-2mm}
\begin{align}\label{eqn:dd_io_i}
C_{IO.i} =& {{K_{i} * b}\over{p}} + {{R}\over{P+1}} - R_{H.i}
\end{align}


This cost is summed for all spilled partitions. The number of spilled partitions, $P_s$, can be estimated by the following inequality (inspired by Formula~\ref{equ:hybridHashParts}), where the remaining $P + 1 - P_{s}$ partitions have enough memory to be completely aggregated in memory:

\vspace{-2mm}
\begin{align}\label{eqn:dd_ps}
{{G}\over{P + 1}} \leq& {{K * (M - P_{s})}\over{M(P + 1 - P_{s})}}
\end{align}


\begin{center}
\begin{table*}[t]
\hspace*{0.3in}
\begin{tabular}{|c||c|c|c|c|c|c|}
\hline
Subsection & Cardinality & Memory & Distribution & HT Slots & Fudge & HH Error \\
\hline\hline
\ref{sec:eval:cmvalid}, \ref{sec:eval:mem}, \ref{sec:eval:card} & 100\%, 44.1\%, & 0.5M $\sim$ 4G & Uniform & 1 & 1.2 & 1  \\
& 6.25\%, 0.02\% & & & & &\\
\hline
\ref{sec:eval:pipe} & 6.25\% & 1M, 64M, 4G & Uniform & 1 & 1.2 & 1 \\
\hline
\ref{sec:eval:hhinput} & 0.02\% & 4M, 16M & Uniform & 1 & 1.2 & $4096 \sim 1/4096$ \\
\hline
\ref{sec:eval:skew} & 1\% & 2M $\sim$ 128M & Uniform, Zipfian, & 1 & 1.2 & 1 \\
& & & Self-Similar, & & & \\
& & & Heavy-Hitter, & & & \\
& & & Sorted & & &\\
\hline
\ref{sec:eval:hash} &&&&&&\\
(hash table slot) & 6.25\% &2M, 4G & Uniform & 1, 2, 3 & 1.2 & 1 \\
\hline
\ref{sec:eval:hash} &&&&&&\\
(fudge factor) & 6.25\% & 2M, 4G & Uniform & 1 & 1.0 $\sim$ 1.6 & 1 \\
\hline
\end{tabular}
\caption{Performance related factors used in the experimental evaluation.}
\end{table*}\label{tbl:exprfactors}
\end{center}

\subsubsection{Pre-Partitioning}
\label{sec:theo:hybridhash:prepartitioning}%
The Pre-Partitioning algorithm 
aggregates records from partition 0 only in its hash-partition phase, while the other $P$ partitions are directly spilled using $P$ output buffers. When bloom filters are used with the hash table slot headers, there is an overhead of one byte per slot, or formally $o' = o + \frac{1}{b}$. The capacity of the list storage area is thus $K = \frac{(M-P)p}{ob + 1}$. Since the algorithm guarantees that partition 0 can be fully aggregated in the hash table, the number of raw input records of partition 0 is $\frac{K}{G_{t}} * R_{t}$. The I/O cost consists of loading the records to be processed and spilling the raw records in the $P$ spilled partitions, i.e.: 
\vspace{-2mm}
\begin{align}
C_{IO} = R + (R - \frac{K}{G_{t}} * R)
\end{align}
The CPU comparison cost includes the cost of hashing the records of partition 0 into the hash table, plus the cost for checking whether a record should be spilled (for records from the $P$ spilling partitions). Assume that the per-slot bloom-filter has a false positive ratio $\alpha$. Then for each of the $(R_{t}*(1 - \frac{K}{G_{t}})$ spilled records, if the bloom filter can detect that the record is not in the hash table, the record is directly flushed (we omit the bloom filter lookup cost since it is negligible compared with the hash comparison cost). If the bloom filter fails to detect that the record is not in the hash table (false positive error with probability $\alpha$), a hash table lookup for the record will cause a hash miss with cost of $\frac{K}{H}$. Therefore the CPU cost is given by:

\begin{align}
C_{comp} =& C_{hash}(\frac{K}{G_{t}} * R_{t}, K, K, H) \nonumber \\
&+ \alpha (R_{t}*(1 - \frac{K}{G_{t}}) * \frac{K}{H} )
\end{align}

\section{Experimental Evaluation}
\label{sec:eval}


We have implemented all algorithms as operators in the Hyracks platform \cite{DBLP:conf/icde/BorkarCGOV11} and performed extensive experimentation. The machine hosting Hyracks is an Intel Xeon E5520 CPU with 16GB main memory and four 10000 rpm SATA disks. We used the Java 6 software environment on 64-bit Linux with kernel version 2.6.18-194.el5. We ran the example query of Section \ref{sec:proenv:func} on a synthetic UserVisits dataset (table) that has two fields: a string \verb|ip| field as the grouping key, containing an abbreviated IPv6 address (from 0000:0001::2001 to 3b9a:ca00::2001 for 1 billion records), and a double \verb|adRevenue| field (randomly generated in $[1, 1000]$). To fully study the algorithm performance and validate the cost models, we consider the variables listed below. The values that we used for these variables in our experiments (organized by subsection) appear in Table \ref{tbl:exprfactors}. 
\vspace{-4mm}
\begin{itemize}
\item {\em Cardinality ratio: } the ratio between the number of raw input records (input size $R$) and the number of unique groups (output size $G$).
\item {\em Memory: } the size of the memory assigned for the aggregation.
\item {\em Data distribution: } the distribution of the groups (keys) in the dataset.
\item {\em Hash table slots: } the number of slots in the hash table, measured by the ratio between the number of slots and the number of unique keys that can fit in the list storage area.
\item {\em Fudge factor: } the hybrid-hash fudge factor.
\item {\em Unique Group Estimation Error: } (applies only to hybrid-hash algorithms) the ratio between the user (query compiler) specified and the actual number of unique groups.
\end{itemize}

\subsection{Cost Model Validation}\label{sec:eval:cmvalid}

To validate the accuracy of our models, we depict the I/O and CPU (as predicted by the models and measured by the experiments) 
of the six algorithms in different memory configurations for two
datasets with cardinality ratios 100\% and
0.02\% in Figures 13 and 14 respectively; we also experimented with
cardinalities 44.1\% and 6.25\% which showed
similar behavior (not shown due to the space limitation).
As we can see, our models can predict both the I/O and the CPU cost with high precision. 
In particular, the I/O cost estimation is consistently very close to the actual I/O. For most cases, the cost for the (hash) CPU comparisons is slightly underestimated by our models because they assume no skew; however, in reality even slightly skewed data will result in higher hash collisions. This explains the slightly lower model prediction for the CPU cost of the hash-based algorithms in Figure 
\ref{fig:model_validate_002}. 

\begin{sidewaysfigure*}
\hspace*{-2.5cm}
\centering$
\begin{array}{cccc}
\psfig{file=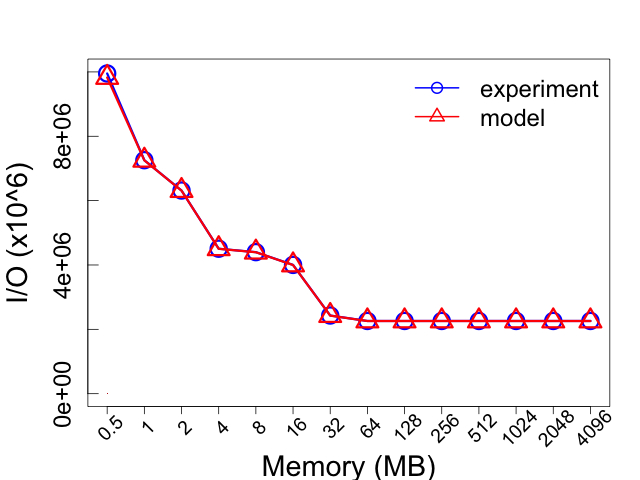,width=2.4in}&
\psfig{file=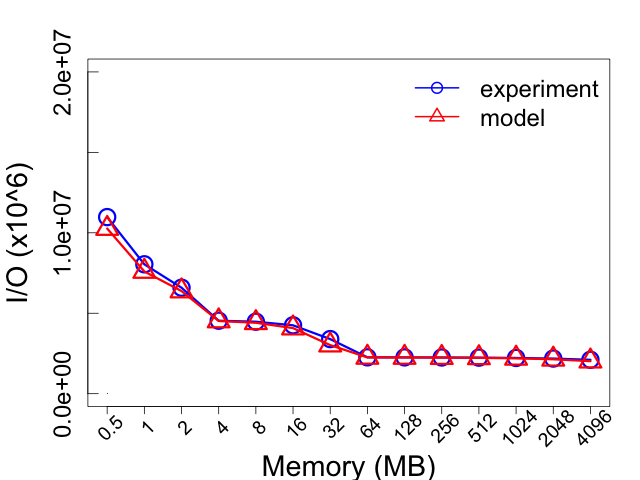,width=2.4in}&
\psfig{file=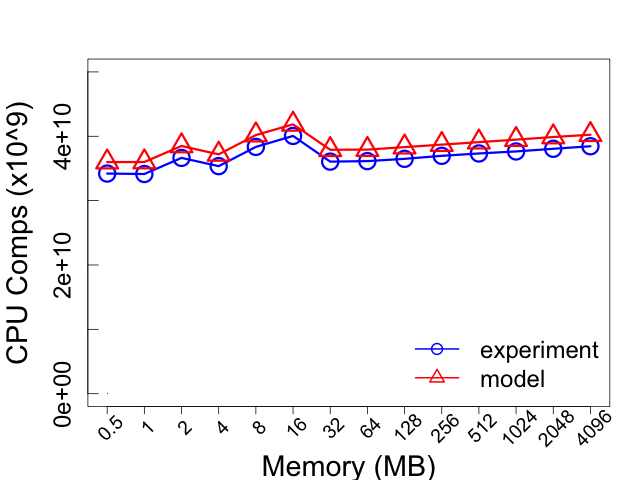,width=2.4in}&
\psfig{file=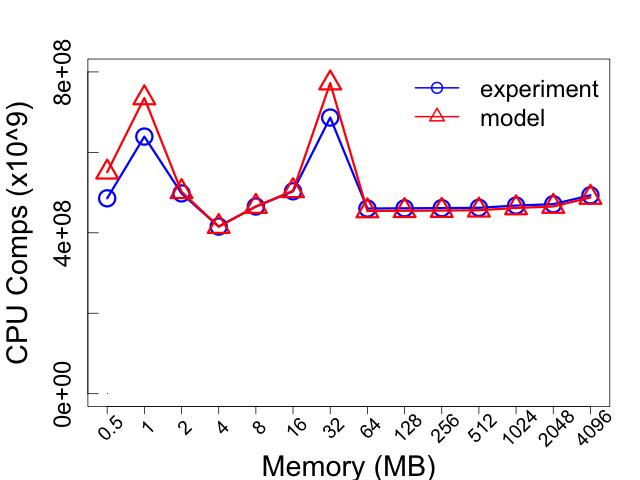,width=2.4in}\\
(a) \mbox{ Sort-based I/O} & (d) \mbox{ Shared Hashing I/O} & (g) \mbox{ Sort-based CPU}  & (j) \mbox{ Shared Hashing CPU} \\
\psfig{file=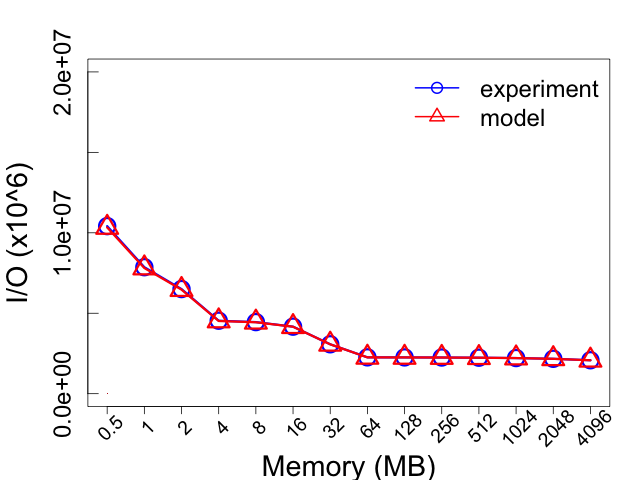,width=2.4in}&
\psfig{file=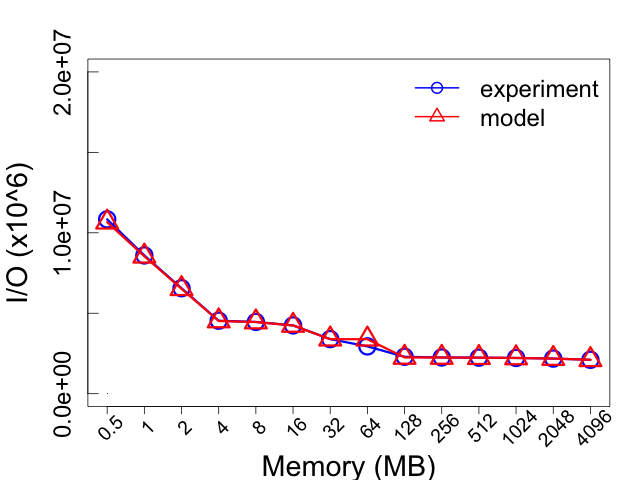,width=2.4in}&
\psfig{file=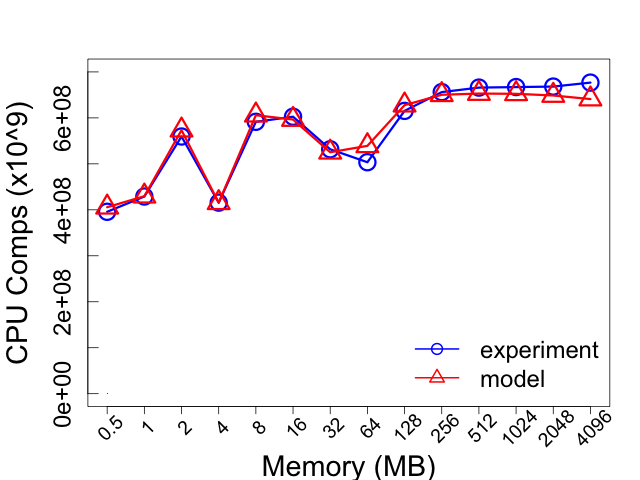,width=2.4in}&
\psfig{file=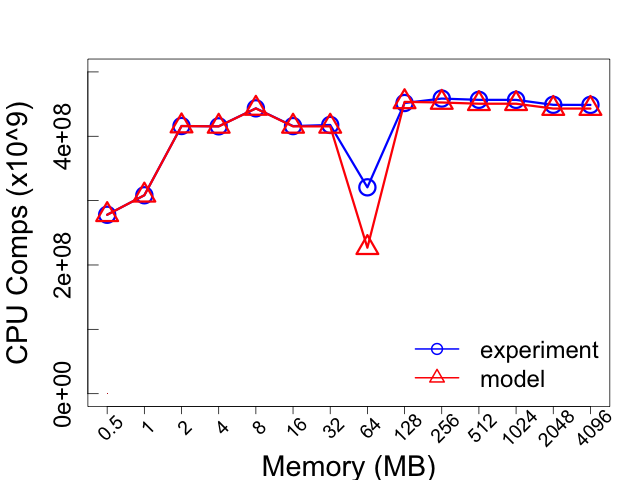,width=2.4in}\\
 (b) \mbox{ Pre-Partitioning I/O} & (e) \mbox{ Original Hybrid-Hash I/O} & (h) \mbox{ Pre-Partitioning CPU} & (k) \mbox{ Original Hybrid-Hash CPU}\\
\psfig{file=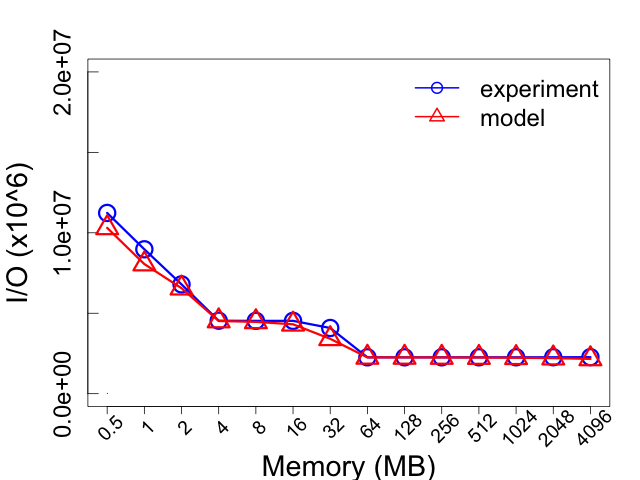,width=2.4in}&
\psfig{file=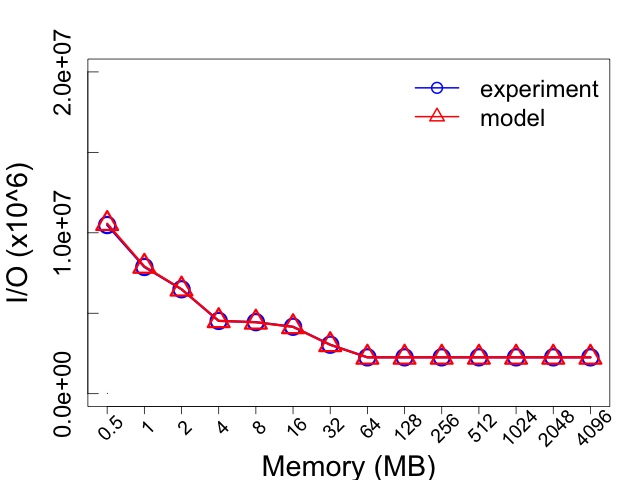,width=2.4in}&
\psfig{file=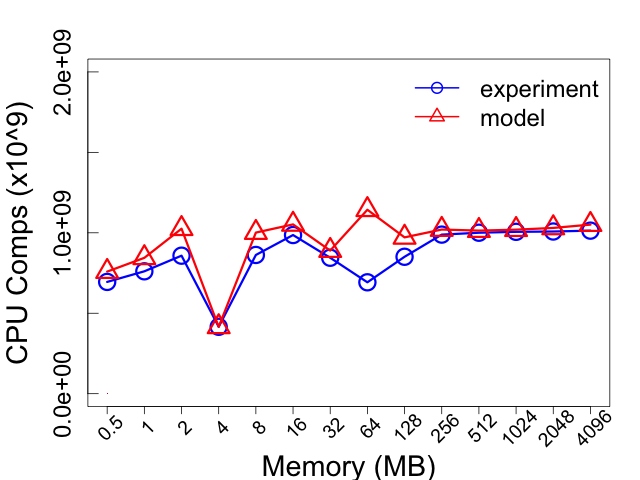,width=2.4in}&
\psfig{file=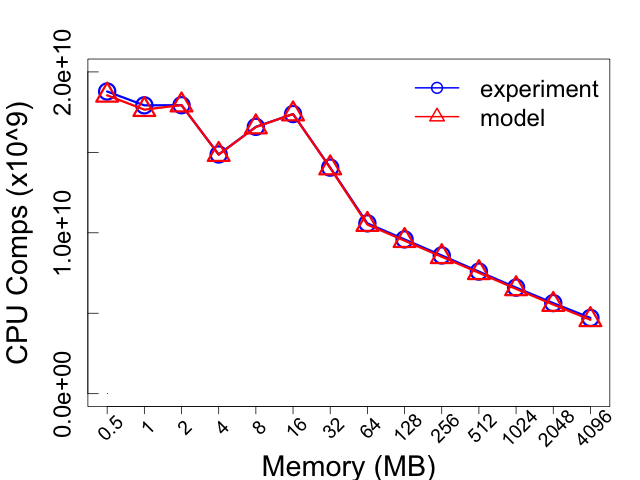,width=2.4in}\\
(c) \mbox{ Dynamic Destaging I/O} & (f) \mbox{ Hash-Sort I/O}   & (i) \mbox{ Dynamic Destaging CPU} & (l) \mbox{ Hash-Sort CPU}\\
\end{array}$
\caption{Model validation (100\% cardinality ratio).}\label{fig:model_validate_100}
\end{sidewaysfigure*}

\begin{sidewaysfigure*}[t]
\centering$
\begin{array}{cccc}
\psfig{file=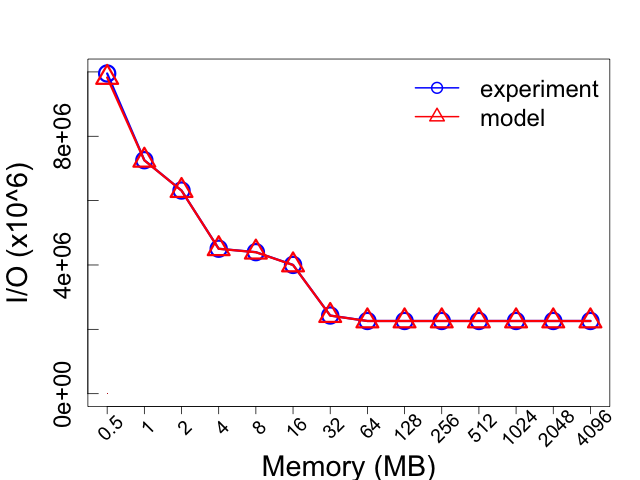,type=png,ext=.png,read=.png,width=2.4in}&
\psfig{file=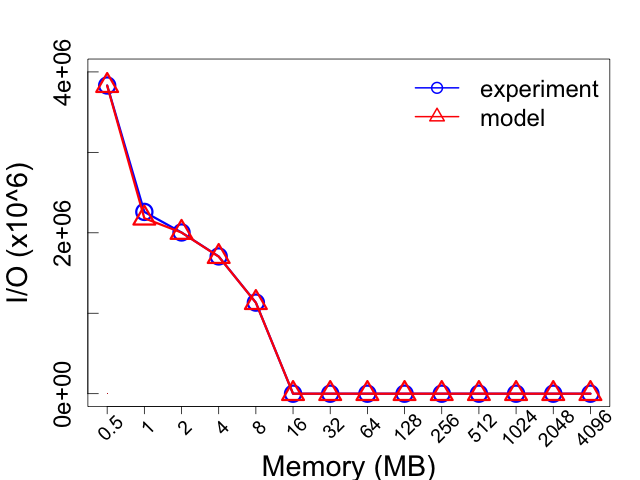,type=png,ext=.png,read=.png,width=2.4in}&
\psfig{file=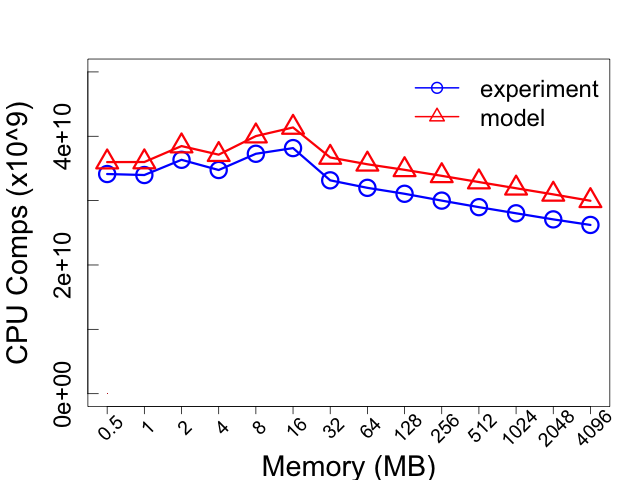,type=png,ext=.png,read=.png,width=2.4in}&
\psfig{file=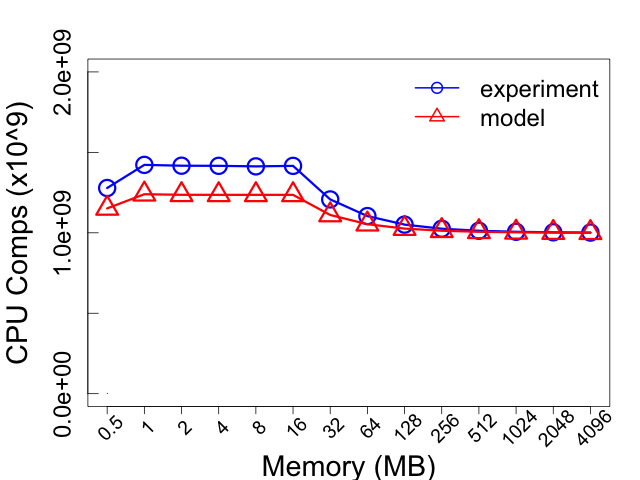,type=png,ext=.png,read=.png,width=2.4in}\\
(a) \mbox{ Sort-based I/O} & (d) \mbox{ Shared Hashing I/O} & (g) \mbox{ Sort-based CPU}  & (j) \mbox{ Shared Hashing CPU} \\
\psfig{file=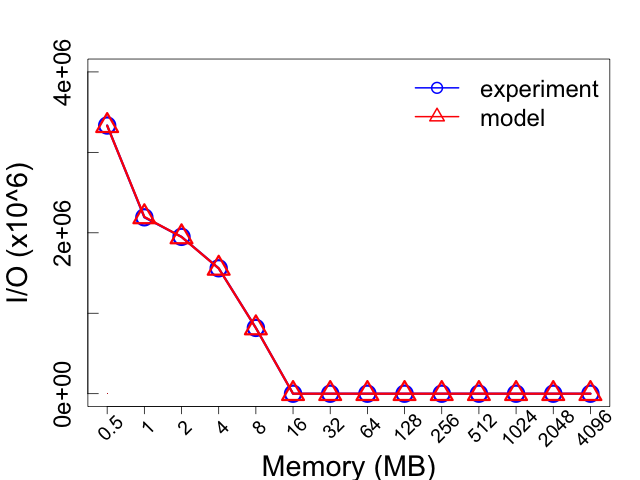,type=png,ext=.png,read=.png,width=2.4in}&
\psfig{file=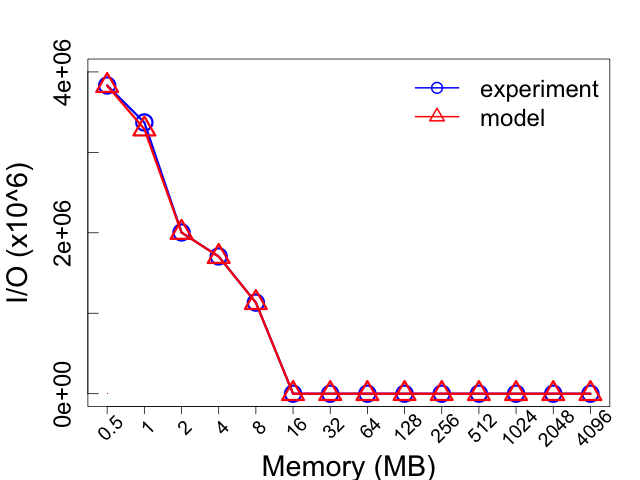,type=png,ext=.png,read=.png,width=2.4in}&
\psfig{file=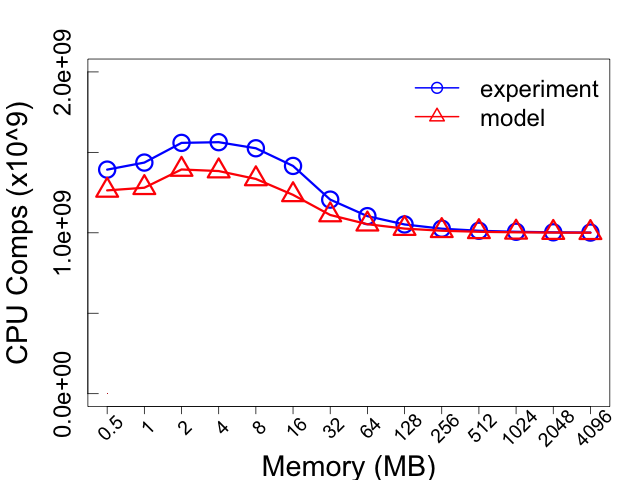,type=png,ext=.png,read=.png,width=2.4in}&
\psfig{file=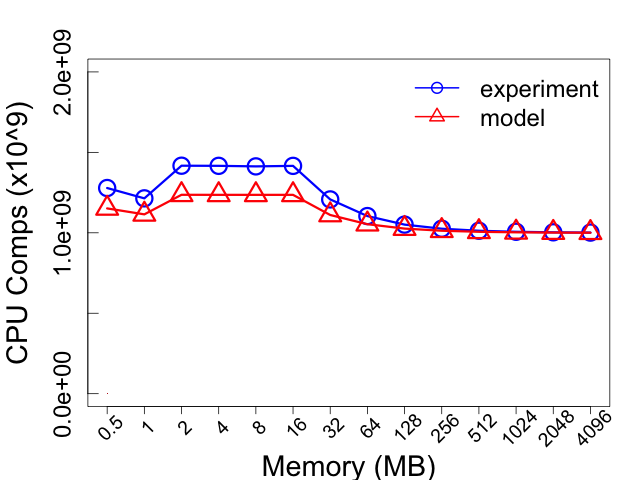,type=png,ext=.png,read=.png,width=2.4in}\\
 (b) \mbox{ Pre-Partitioning I/O} & (e) \mbox{ Original Hybrid-Hash I/O} & (h) \mbox{ Pre-Partitioning CPU} & (k) \mbox{ Original Hybrid-Hash CPU}\\
\psfig{file=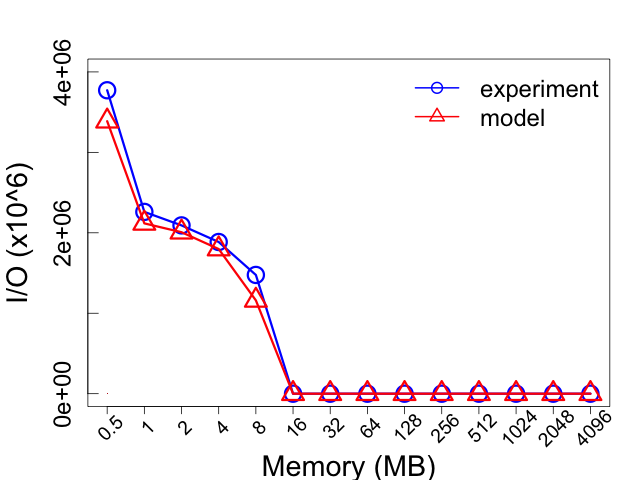,type=png,ext=.png,read=.png,width=2.4in}&
\psfig{file=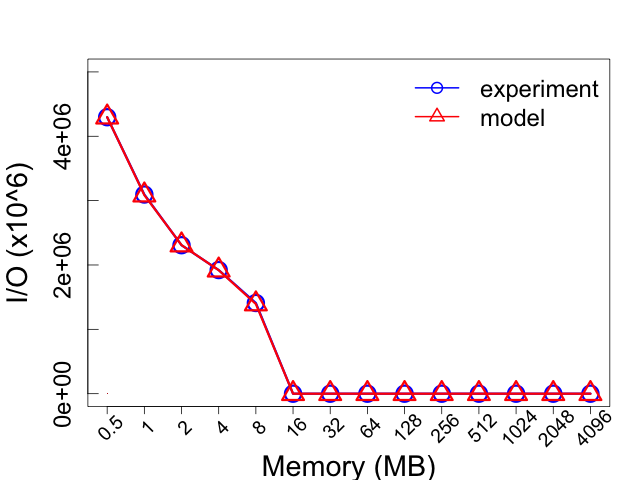,type=png,ext=.png,read=.png,width=2.4in}&
\psfig{file=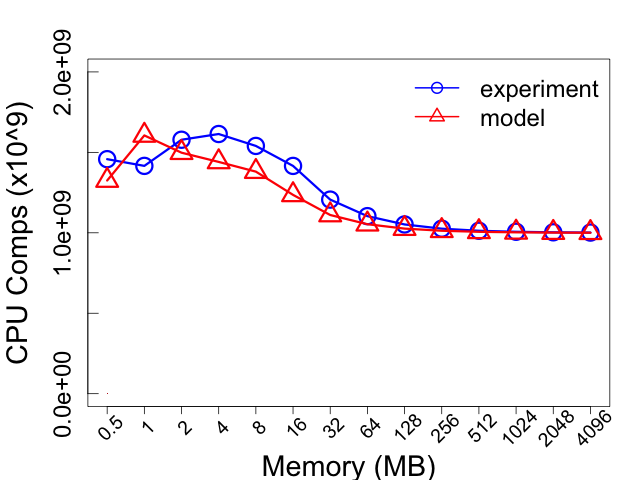,type=png,ext=.png,read=.png,width=2.4in}&
\psfig{file=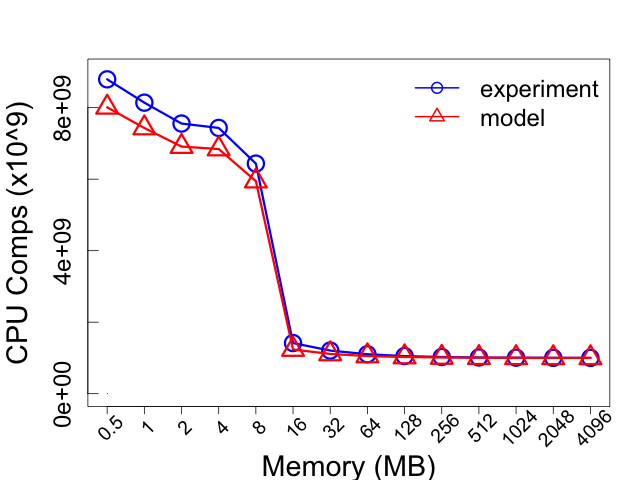,type=png,ext=.png,read=.png,width=2.4in}\\
(c) \mbox{ Dynamic Destaging I/O} & (f) \mbox{ Hash-Sort I/O}   & (i) \mbox{ Dynamic Destaging CPU} & (l) \mbox{ Hash-Sort CPU}\\
\end{array}$
\caption{Model validation (0.02\% cardinality ratio).}\label{fig:model_validate_002}
\end{sidewaysfigure*}

There are cases where our models overestimate the CPU cost, as when processing the ``all unique'' dataset (Figure \ref{fig:model_validate_100}, with cardinality ratio 100\%) for the Dynamic Destaging and Shared Hashing algorithms. This is because with actual data, the hash table spilling could be triggered earlier than the model prediction since the key distribution is not perfectly uniform; as a result, less groups from spilling partitions are hashed into the dynamic/shared hash scheme, leading to less actual CPU cost. 

Among all algorithms, the CPU model for Dynamic Destaging showed the largest overestimation compared to the real experiments for some configurations. The reason is that in these cases, our cost model assumes that the resident partition can be completely aggregated in-memory, however in reality our experiments show that in these configurations, the resident partition has also been spilled due to the imperfect hash partitioning and dynamic destaging (i.e., evicting the right partitions for spilling) in reality. When reloading the spilled resident partition, the number of hash table collisions is less since the records are hashed to the whole hash table space instead of just a potion of it, so the actual CPU comparison cost is less than predicted. 

\subsection{Effect of Memory Size}\label{sec:eval:mem}
\label{sec-8-2-1}

\begin{sidewaysfigure*}
\hspace*{-2.5cm}
\centering$
\begin{array}{cccc}
\multicolumn{4}{c}{\hspace*{-0.2in}\psfig{file=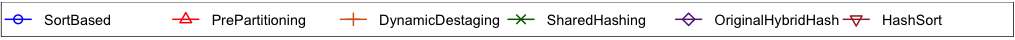, height=0.3in}}\\
\psfig{file=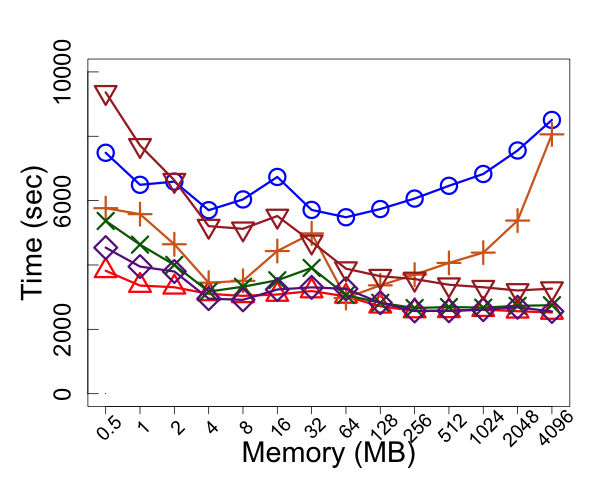,type=png,ext=.png,read=.png,width=2.4in}&
\psfig{file=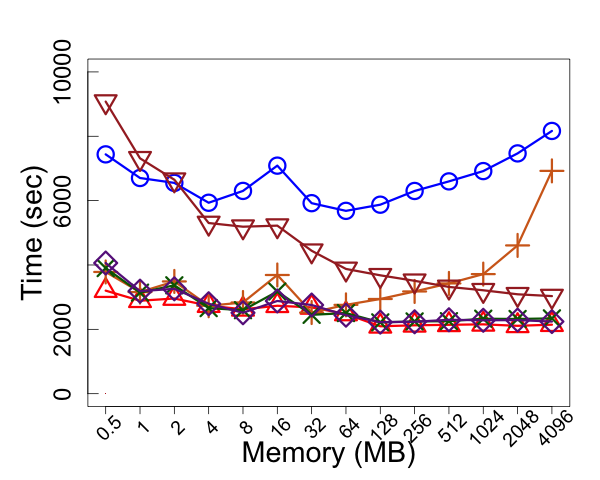,type=png,ext=.png,read=.png,width=2.4in}&
\psfig{file=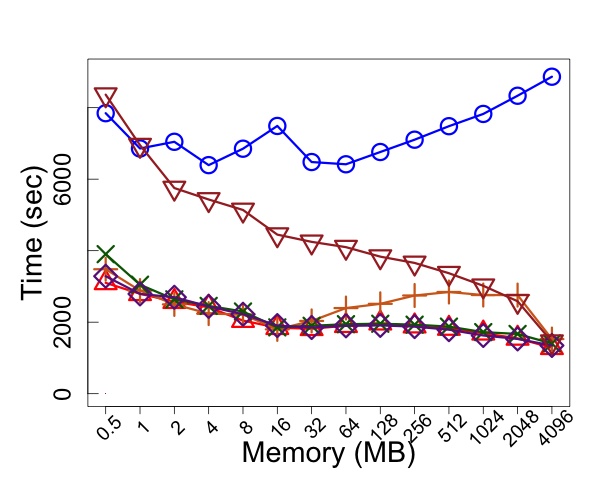,type=png,ext=.png,read=.png,width=2.4in}&
\psfig{file=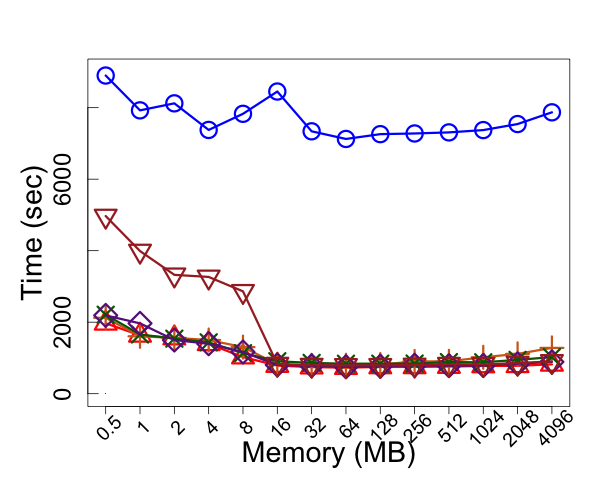,type=png,ext=.png,read=.png,width=2.4in}\\
(a)\mbox{ Running Time (100\%)} & (b)\mbox{ Running Time (44.1\%)} & (c) \mbox{ Running Time (6.25\%) } & (d) \mbox{ Running Time (0.02\%) }\\
\psfig{file=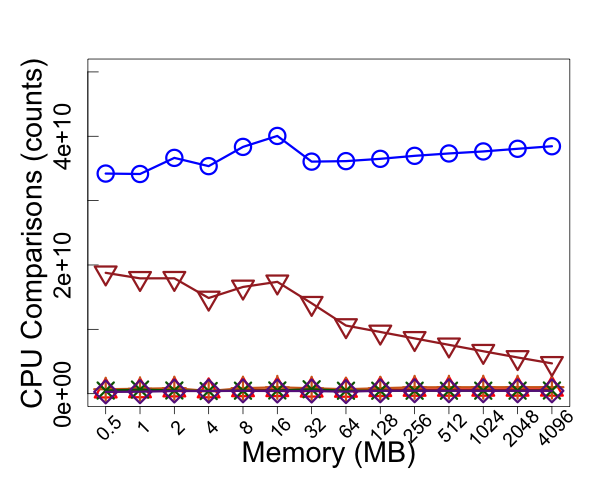,type=png,ext=.png,read=.png,width=2.4in}&
\psfig{file=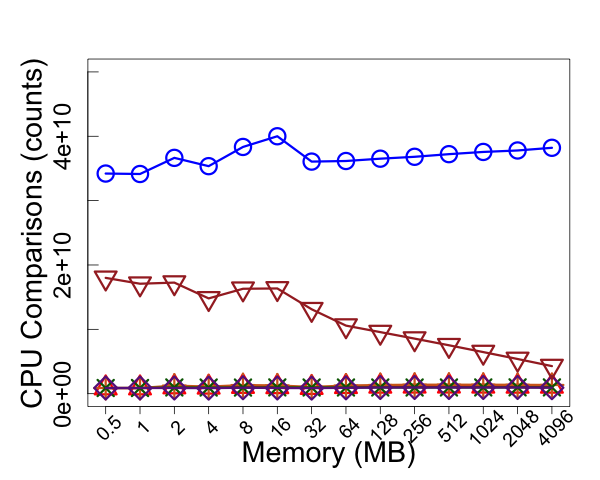,type=png,ext=.png,read=.png,width=2.4in}&
\psfig{file=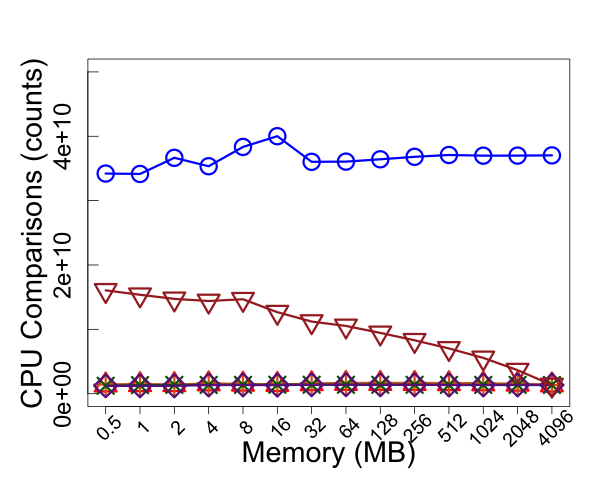,type=png,ext=.png,read=.png,width=2.4in}&
\psfig{file=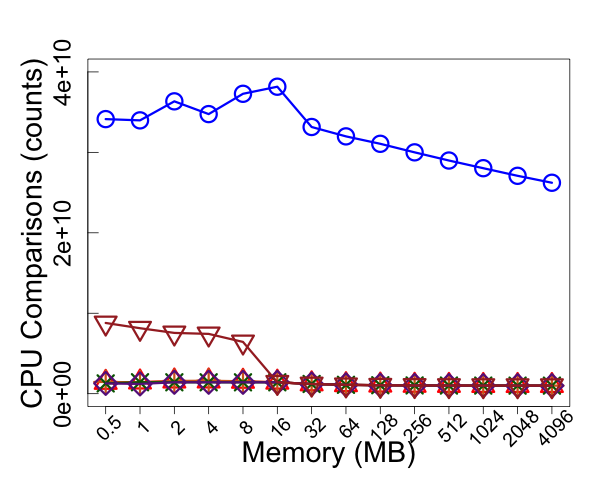,type=png,ext=.png,read=.png,width=2.4in}\\
(e)\mbox{ CPU Comparisons (100\%)} & (f)\mbox{ CPU Comparisons (44.1\%)} & (g) \mbox{ CPU Comparisons (6.25\%) } & (h) \mbox{ CPU Comparisons (0.02\%) }\\
\psfig{file=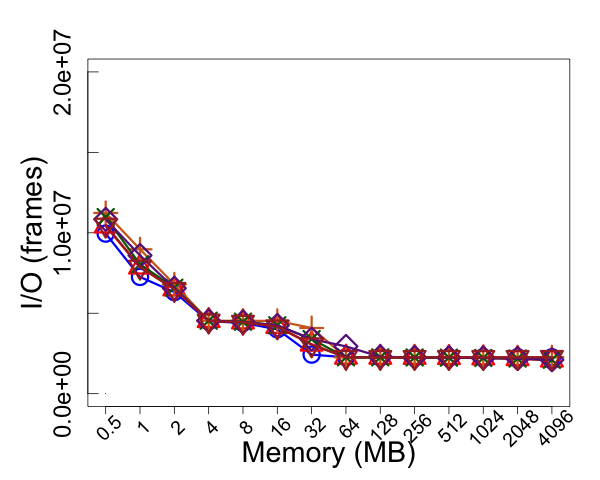,type=png,ext=.png,read=.png,width=2.4in}&
\psfig{file=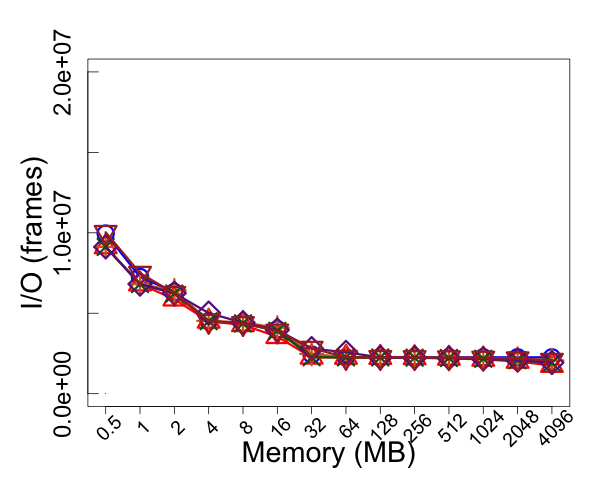,type=png,ext=.png,read=.png,width=2.4in}&
\psfig{file=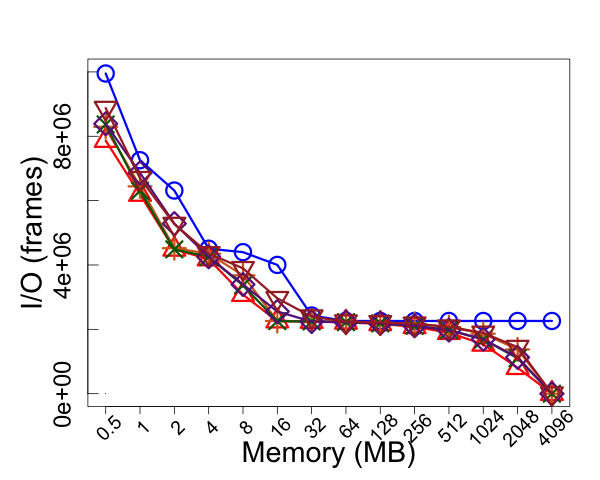,type=png,ext=.png,read=.png,width=2.4in}&
\psfig{file=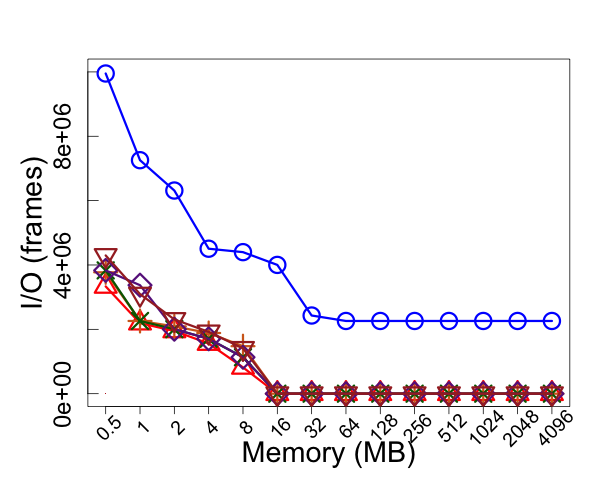,type=png,ext=.png,read=.png,width=2.4in}\\
(i)\mbox{ I/O (100\%)} & (j)\mbox{ I/O (44.1\%)} & (k) \mbox{ I/O (6.25\%) } & (l) \mbox{ I/O (0.02\%) }
\end{array}$
\caption{Experiments with different cardinality ratios and memory sizes.}\label{fig:cards}
\end{sidewaysfigure*}

To study the effect of memory size on the aggregation algorithms we measured their running time using the four uniform data sets (with cardinality ratio 100\%, 44.1\%, 6.25\% and 0.02\%) in different memory configurations (0.5M to 4G). (The effects of skewed data are examined later). In Figure~\ref{fig:cards}, we show the running time, CPU comparison cost and I/O cost for all these experiments. 
When considering the CPU cost, the algorithms that use sorting require more CPU than the pure hashing algorithms. The I/O cost for all algorithms decreases when memory increases (since more records can be aggregated in memory).

We first observe that for larger memories (memory larger than 64M) the running time of the Sort-based algorithm increases. This is because larger memory settings cause higher cache misses for the comparing and swapping in the sorting procedure. Furthermore, when the cardinality ratio is high (100\%, 44.1\%, and 6.25\%), the total CPU cost for sorting is increasing according to Formula~\ref{equ:sort} of the sort component in Appendix~\ref{sec-6-2}(the records to be sorted in each full memory chunk $m$ is larger). This can also be observed through the similar rising of the CPU cost for memories larger than 64M (Figure \ref{fig:cards} (e-h)). 
Thus it is not always the case that larger memory leads to better performance in the Sort-based algorithm. Different from the Sort-based algorithm, the Hash-Sort algorithm has better performance when the memory is larger because it utilizes collapsing, and most of the time it is faster than the Sort-based algorithm (except for the case with small memory, where the collapsing cannot be fully exploited). 

The four hybrid-hash algorithms have the best performance since they avoid sorting and merging. 
Among the hybrid-hash algorithms, the Pre-Partitioning algorithm has the most robust performance along all memory and key cardinality configurations. This is because Pre-Partitioning always creates the resident partition to fill up the in-memory hash table. This will reduce both the I/O (since more groups are aggregated within the resident partition) and the CPU comparison cost (since less spilled records need to be processed recursively). Furthermore, by using bloom filters within the hash table, the extra cost for hash misses is reduced so its CPU cost is just slightly higher than the Original Hybrid-Hash algorithm (as showed in Figure~\ref{fig:prepartbf}). 

Also note that according to Formula~\ref{equ:hybridHashParts}, the memory space reserved for the resident partition $(M - P) = {{M^2 - 3M - G * F}\over{M - 2}}$ is not linearly associated with the memory size. This means that when the memory increases, although the number of hash table slots increases correspondingly, the size of the resident partition does not increase linearly. So the hash collision could vary based on the ratio between the unique records in the resident partition and the hash table slots. In the case that this ratio is higher due to a larger increase of the unique records in the resident partition than the increase of the hash table slots, there will be more hash comparison cost for aggregating the resident partition. 
This explains the spikes of the CPU cost for all hybrid-hash algorithm along different memory configurations. 

We further notice that the running time for Dynamic Destaging is increasing (it becomes larger than the other hybrid-hash algorithms) for memories between 16M and 2048M. In these memory configurations only one round of hybrid-hash is needed (i.e., no grace partition is used). However the partition tuning optimization \cite{DBLP:conf/vldb/Graefe99} increases the number of partitions as the memory increases, which causes more cost overhead for maintaining the spilling files. Furthermore, as the memory increases, the number of records from spilling partitions that have been partially aggregated and flushed will be larger (recall that in Dynamic Destaging, spilling partitions are dynamically spilled in order to maximize the in-memory aggregation); this could potentially increase the hashing cost because all partial results must be reloaded and hashed again. 

Finally when the memory size is relatively very large (4G), all hybrid-hash algorithms have the same running time, as no spilling happens (so all can do in-memory aggregation). 

\clearpage

\begin{sidewaysfigure*}[t]
\centering$
\begin{array}{ccccc}
\multicolumn{5}{c}{\hspace*{-0.2in}\psfig{file=legends.png, height=0.3in}}\\
\psfig{file=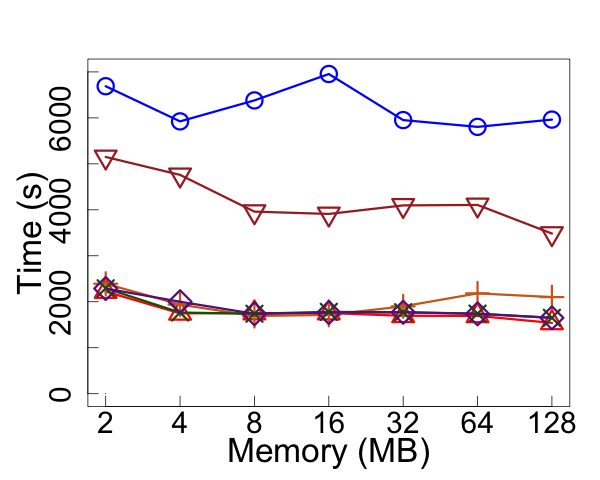,type=png,ext=.png,read=.png,width=2in}&
\psfig{file=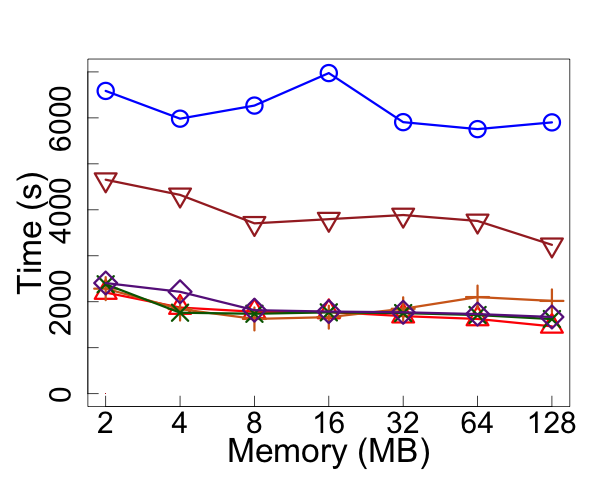,type=png,ext=.png,read=.png,width=2in}&
\psfig{file=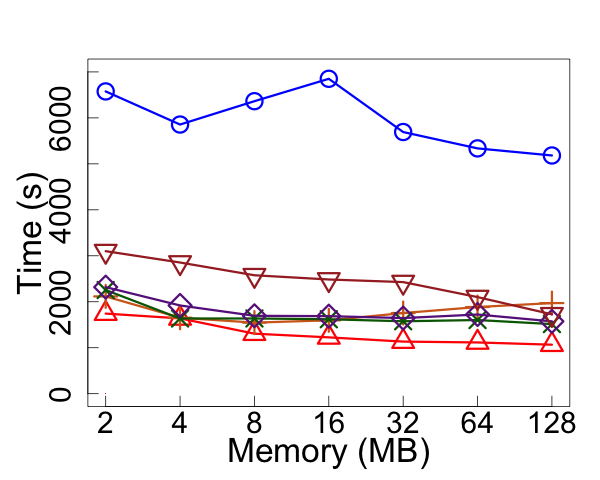,type=png,ext=.png,read=.png,width=2in}&
\psfig{file=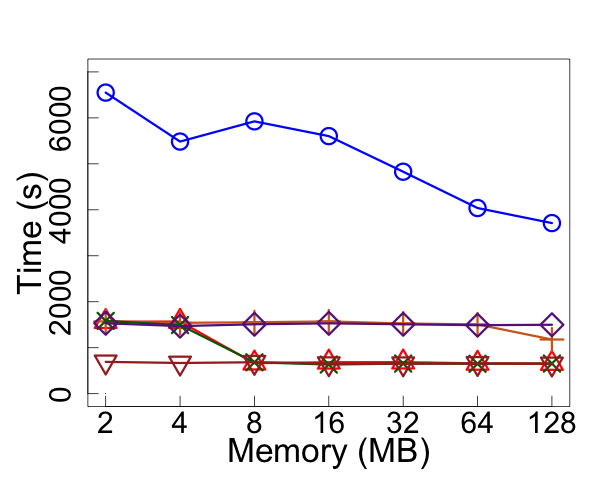,type=png,ext=.png,read=.png,width=2in}&
\psfig{file=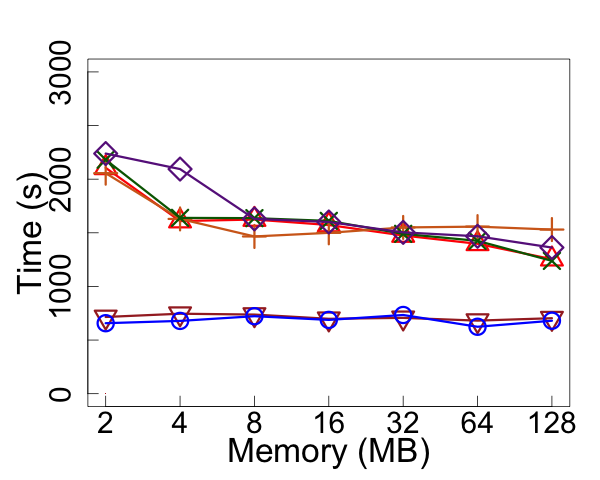,type=png,ext=.png,read=.png,width=2in}\\
(a) \mbox{ Time (Uniform)} & (b) \mbox{ Time (Zipfian 0.5)} & (c) \mbox{ Time (Self-Similar 20-80)}  & (d) \mbox{ Time (Heavy-Hitter)} & (e) \mbox{ Time (Uniform-Sorted)} \\
\psfig{file=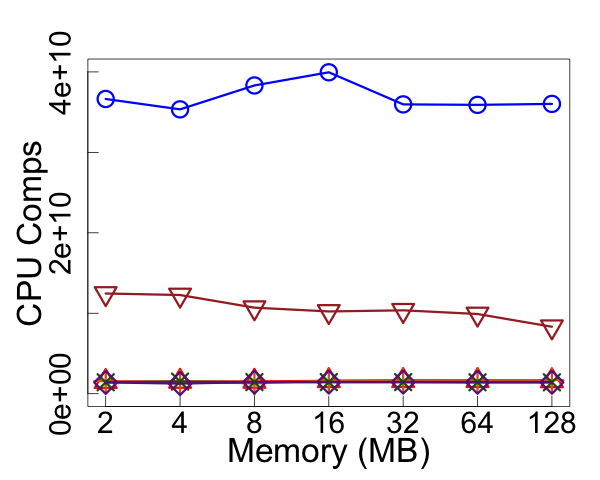,type=png,ext=.png,read=.png,width=2in}&
\psfig{file=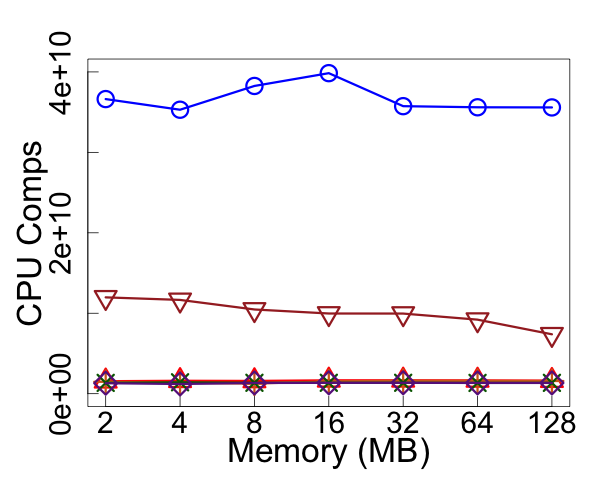,type=png,ext=.png,read=.png,width=2in}&
\psfig{file=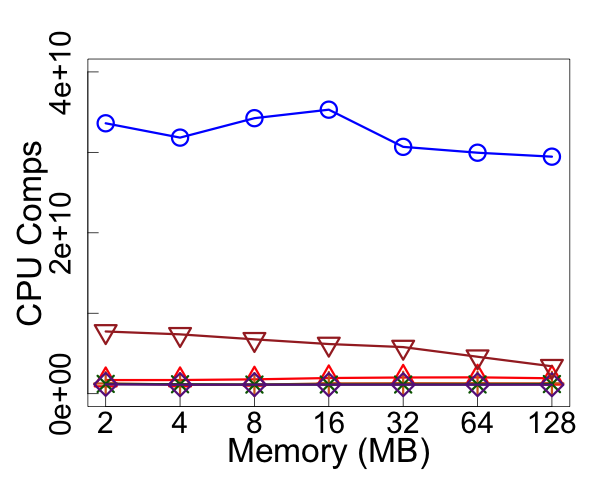,type=png,ext=.png,read=.png,width=2in}&
\psfig{file=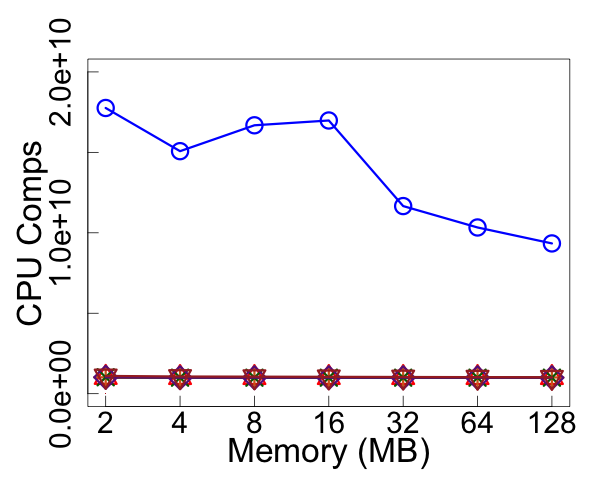,type=png,ext=.png,read=.png,width=2in}&
\psfig{file=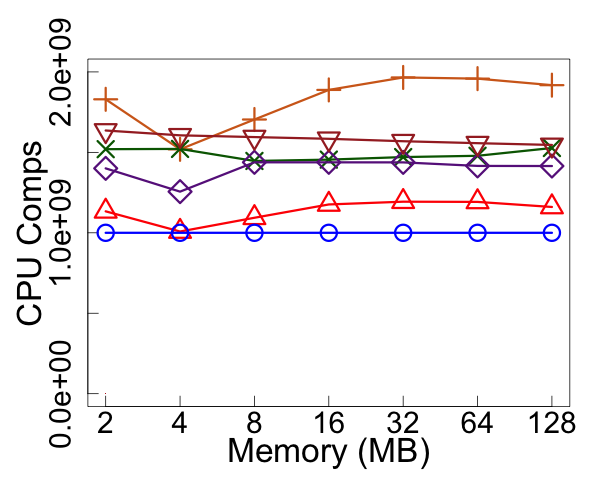,type=png,ext=.png,read=.png,width=2in}\\
(f) \mbox{ CPU (Uniform)} & (g) \mbox{ CPU (Zipfian 0.5)} & (h) \mbox{ CPU (Self-Similar 20-80)}  & (i) \mbox{ CPU (Heavy-Hitter)} & (j) \mbox{ CPU (Uniform-Sorted)} \\
\psfig{file=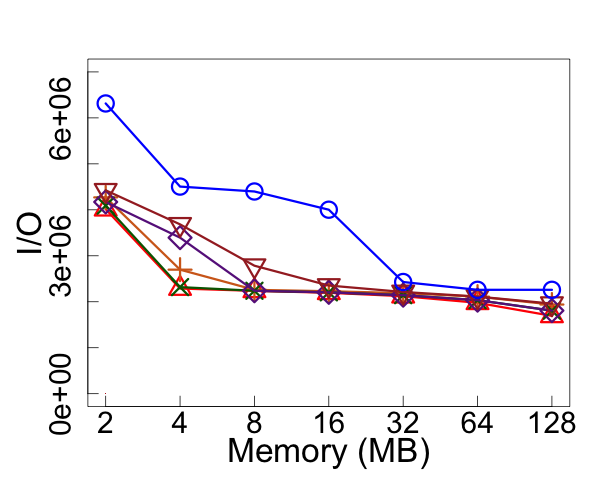,type=png,ext=.png,read=.png,width=2in}&
\psfig{file=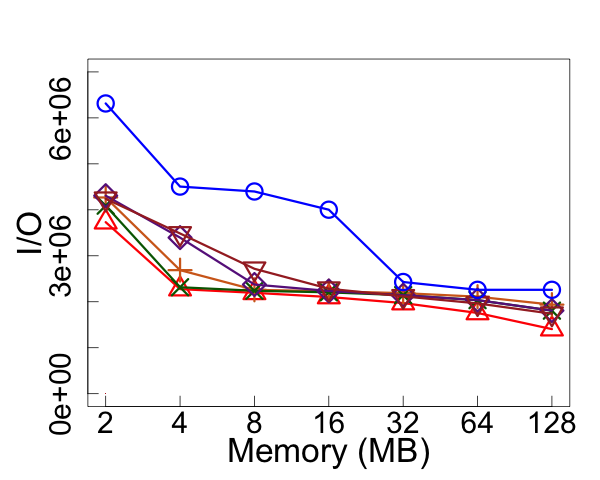,type=png,ext=.png,read=.png,width=2in}&
\psfig{file=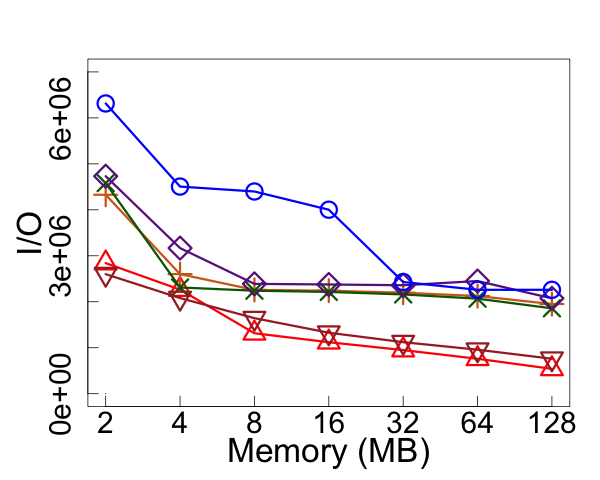,type=png,ext=.png,read=.png,width=2in}&
\psfig{file=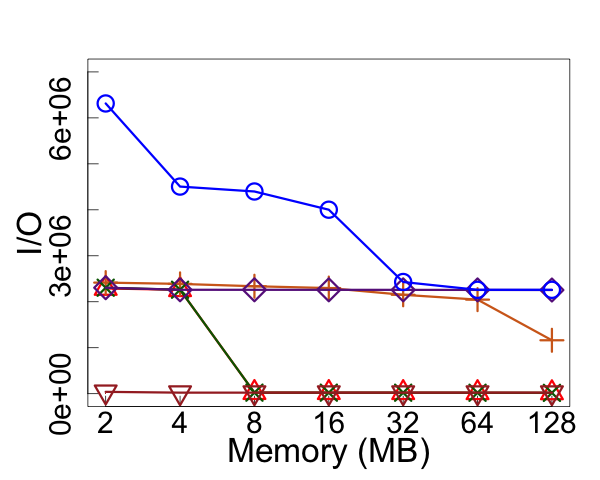,type=png,ext=.png,read=.png,width=2in}&
\psfig{file=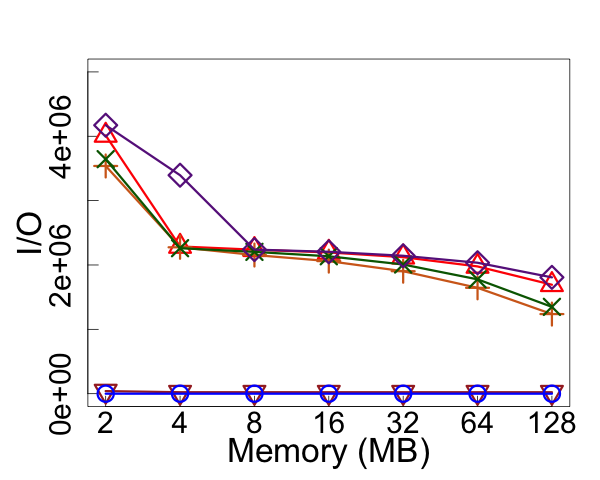,type=png,ext=.png,read=.png,width=2in}\\
(k) \mbox{ I/O (Uniform)} & (l) \mbox{ I/O (Zipfian 0.5)} & (m) \mbox{ I/O (Self-Similar 20-80)}  & (n) \mbox{ I/O (Heavy-Hitter)} & (o) \mbox{ I/O (Uniform-Sorted)}
\end{array}$
\caption{Experiments on skew datasets.}\label{fig:skews}
\end{sidewaysfigure*}
\clearpage

\subsection{Effect of Cardinality Ratio}\label{sec:eval:card}



Figure~\ref{fig:cards} also compares the algorithms for different cardinality ratios. Note that full in-memory aggregation happens for the 0.02\% dataset when memory is larger than 16M, while it occurs for the 6.25\% dataset only for the 4G memory. In the higher cardinality ratio datasets 
(100\% and 44.1\%) there is no in-memory aggregation (since the number of grouping keys is so large that all algorithms need to spill).

The Sort-based algorithm is typically slower than the rest, with the hybrid-hash algorithms being the fastest and the Hash-Sort falling in between (except for the case of very small memories and high cardinalities to be discussed below). As the cardinality ratio increases, the gap in performance between the Sort-based and the hybrid-hash algorithms is reduced. This is because a higher cardinality means more unique keys, and thus less collapsing, which reduces the advantage of hashing. 
Note that when the memory is small and the cardinality is large, the Hash-Sort algorithm performs even worse than the Sort-based algorithm because there is very limited benefit from early aggregation and the hash cost is almost wasted. 

The hybrid-hash-based algorithms are greatly affected by the higher cardinality ratio, as fewer records can be collapsed through aggregation and the performance mainly depends on the effectiveness of partitioning. The spikes in the CPU cost (caused by the non-linear correlation between the resident partition size and the hash table size; see the discussion in the previous subsection) are more clear for data sets with higher cardinality ratios since the hash miss cost is more significant. 




\subsection{Aggregating Skewed Data}\label{sec:eval:skew}

To examine the performance of the algorithms when aggregating skewed data, we considered the following skewed datasets, each with 1 billion records and 10 million unique keys, generated using the algorithms described in \cite{DBLP:conf/sigmod/GraySEBW94}:
\begin{itemize}
\item {\em Uniform}: all unique keys are uniformly distributed among the input records;
\item {\em Zipfian}: we use skew parameter 0.5;
\item {\em Self-similar}: we use the 80-20 proportion;
\item {\em Heavy-hitter}: we choose one key to have $10^9 - (10^7 - 1)$ records, while all other keys have only one record each;
\item {\em Sorted uniform}: we use a uniform data set with records sorted on the grouping key.
\end{itemize}


Figure~\ref{fig:skews} shows the running time, CPU cost and I/O cost of all algorithms for different skew distributions.
Overall we observe that if the skew distribution is similar to the uniform distribution (the Zipf and the Self-Similar data sets), the behaviors of the algorithms are similar to the uniform case. A common characteristic of the two less-skewed datasets (Zipf and Self-Similar) is that the duplicates are distributed in a ``long-tail'' pattern. There are a few keys with very many duplicates (the peak of the distribution) and many keys with very few duplicates (the tail part). Nevertheless, statistically the peak in the Zipf dataset is lower than the peak in the Self-Similar dataset and its long-tail part is higher than the long tail of the Self-Similar dataset. Since there are more duplicates per key in the Zipf dataset, more hash comparisons are needed. 

\begin{figure*}[t!]
\centering$\begin{array}{ccc}
\psfig{file=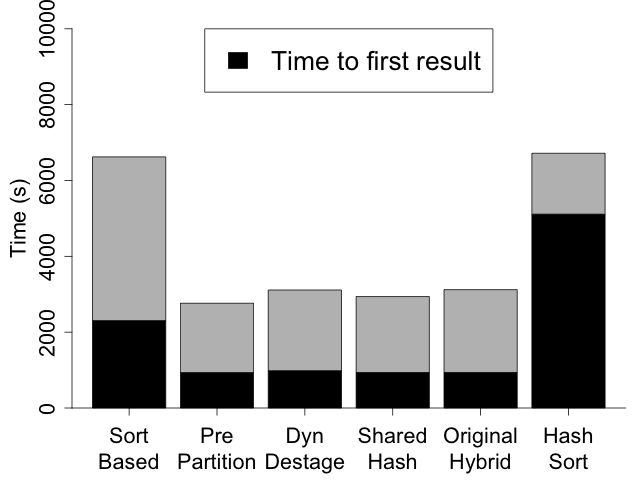, width=2.1in} &
\psfig{file=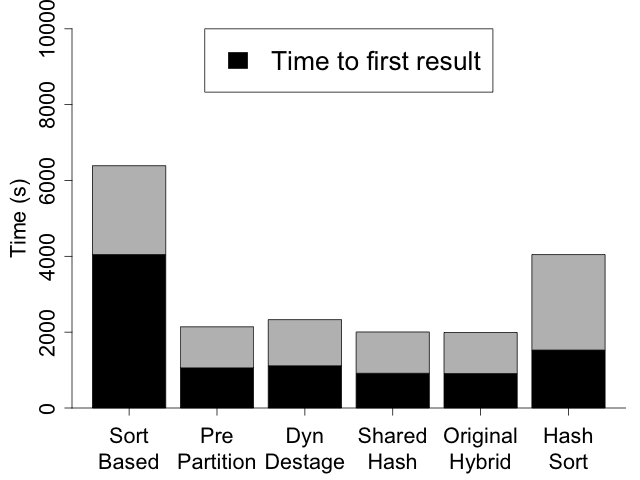, width=2.1in} &
\psfig{file=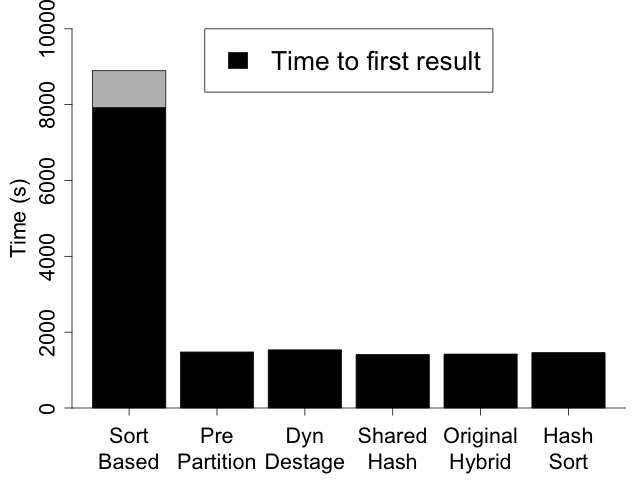,width=2.1in} \\
(a) \mbox{ 1M memory} & (b) \mbox{ 64M memory} & (c) \mbox{ 4G memory}
\end{array}$
\caption{Time to the first result as part of the total running time.}\label{fig:pipeline}
\label{fig:models}
\end{figure*}

For the Zipf and Self-Similar datasets, the Hash-Sort algorithm is overall slower than the hybrid-hash based algorithms because (1) these datasets are not sorted, so the Hash-Sort algorithm needs to sort and merge the intermediate results, and (2) since more grouping keys have duplicates, the same grouping key could be in multiple run files, which further increases the run file size and the cost for merging.
For these datasets, the Pre-Partitioning algorithm has the best running time since it always fills up the memory space reserved for the resident partition, so more groups can be collapsed into the resident partition. This greatly reduces the total I/O cost for the Pre-Partitioning algorithm compared with other hybrid-hash algorithms, leading to a lower running time. 
It is interesting to note that this behavior is more apparent in the Self-Similar than the Zipf dataset. This is because the tail part in the Self-Similar dataset is smaller, so the size of the spilling partitions would be smaller when compared with the Zipf dataset. This will reduce both the hash miss cost for checking the spilling records and the I/O cost for spilling partitions.

For the Heavy-Hitter and the Uniform-Sorted datasets, the nature of their skew is more significant compared with the uniform case, so their behaviors are quite different than the uniform case.
In particular, for the Heavy-Hitter data set, 
the Hash-Sort algorithm has the best overall performance. The algorithm collapses many duplicates in this data set in its early aggregation; moreover, its slot-based sorting strategy can minimize the sorting cost for merging. The Original Hybrid-Hash algorithm performs the worst in this case because the partition containing the heavy hitter key contains 99\% of the total records; this causes the algorithm to fallback to the Hash-Sort (because it has more than 80\% of the original input content as mentioned in Section~\ref{sec:algs:hybridhash}). 
The Dynamic Destaging algorithm also performs bad due to the fallback, but the fallback is triggered by partition tuning. This is because partitions that do not contain the heavy hitter key are underestimated on their grouping key cardinality, and partition tuning merges them based on the underestimated cardinality. After merging is done, the key cardinality is greater than the memory capacity so these partitions are spilled again. Finally all spilled partitions are processed through the fallback algorithm (the hybrid-hash level is deeper than a Sort-based algorithm), resulting in longer running time. 
The Shared Hashing algorithm performs better when grace partitioning is not needed because it collapses the partition containing the heavy hitter by maximizing the in-memory aggregation through the shared hash table. A similar effect happens for the Pre-Partitioning algorithm, but it performs better since it always guarantees that the resident partition can be completely aggregated in memory. 
For the uniform-sorted dataset, the Sort-based algorithm performs the best since it only needs a single scan over the sorted data to finish the aggregation.
The Hash-Sort algorithm still shows good running times because it can aggregate each group completely in the sorted run generation phase, utilizing the sort order. However it is slightly slower than the Sort-based algorithm due to its higher I/O cost (because of the overhead of the hash table) and the CPU cost (since hashing is more expensive than the sequential match-and-scan procedure).
The four hybrid-hash algorithms perform worse because all partitions produced by grace partitioning (for the 2M and 4M memory) or by the hybrid-hash algorithms (for 8M or larger memory) have to be processed by a recursive hybrid-hash procedure. With 4M memory, the Original Hybrid-Hash performs worse than the other hybrid-hash algorithms because they have better hash collapsing effect; as a result, they can finish the hybrid hash aggregation one level earlier than the Original Hybrid-Hash algorithm using less I/O. 

\subsection{Time to First Result (Pipelining)}\label{sec:eval:pipe}

To check whether these algorithms can be pipelined effectively, we measure the time needed to produce the first aggregation result as another aspect of their performance.  Figure~\ref{fig:pipeline} depicts the results using the 6.25\% dataset in three different memory configurations. The full bar height corresponds to the total running time (full aggregation), while the bottom solid part corresponds to the time until the first aggregation result is produced. The earlier the aggregation result is produced, the better the algorithm can fit into a pipelined query stream. 

\renewcommand{\arraystretch}{0.1}

\begin{figure*}
\centering$\begin{array}{ccc}
\multicolumn{3}{c}{\psfig{file=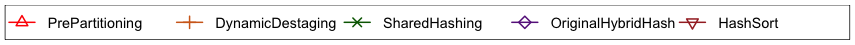, height=0.25in}}\\
\psfig{file=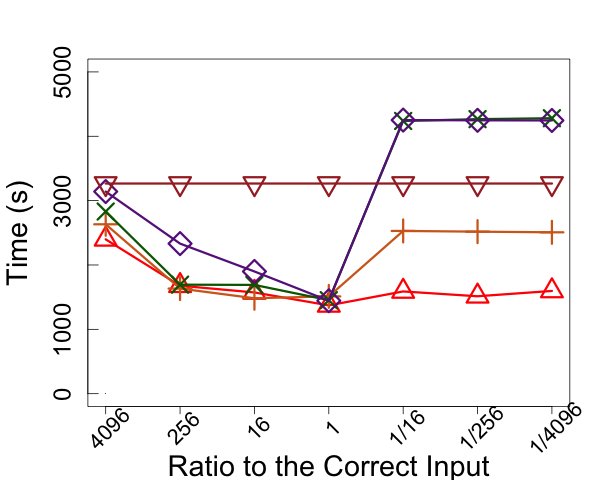, width=1.9in} &
\psfig{file=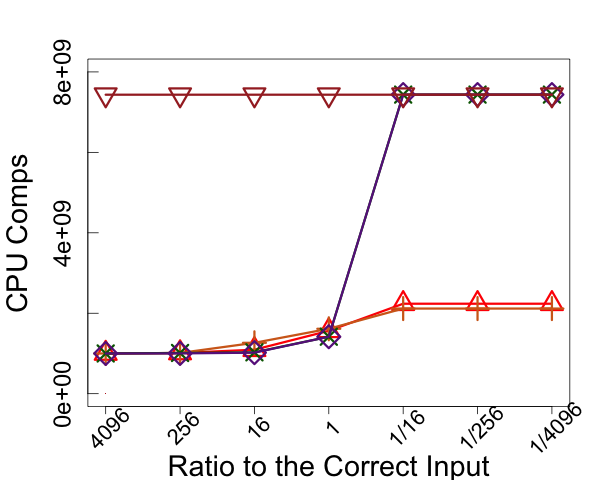, width=1.9in} &
\psfig{file=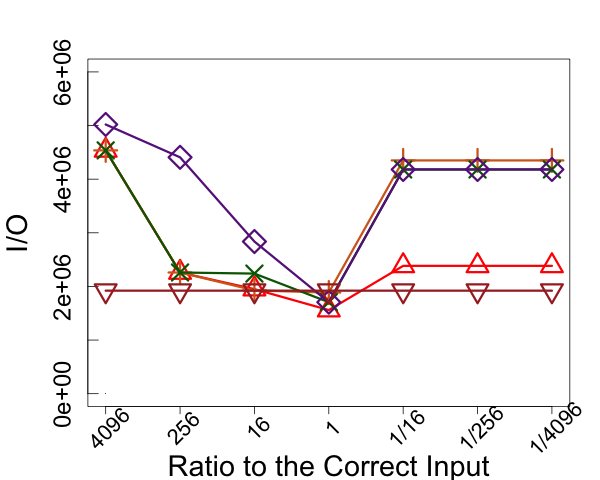,width=1.9in} \\
&&\\
&&\\
(a) \mbox{ Runing Time (4M)} & (b) \mbox{ CPU Comparisons (4M)} & (c) \mbox{ I/O (4M)}\\
\psfig{file=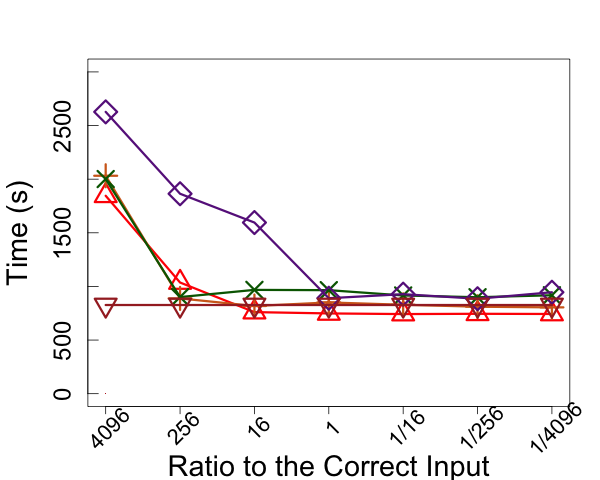, width=1.9in} &
\psfig{file=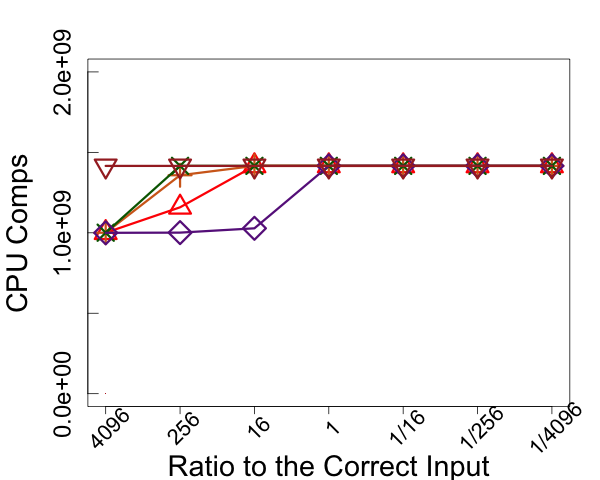, width=1.9in} &
\psfig{file=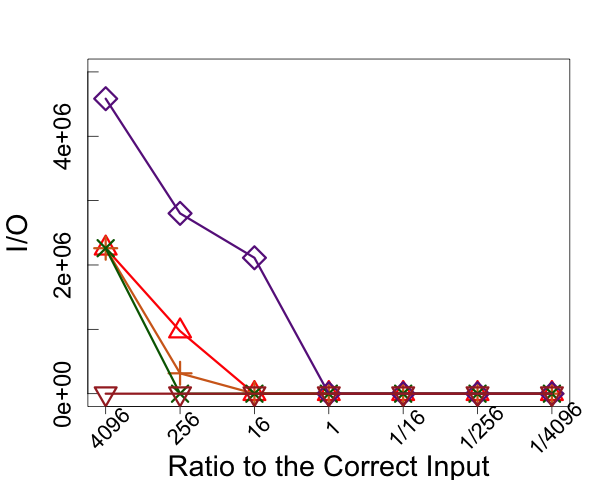,width=1.9in} \\
&&\\
&&\\
(d) \mbox{ Runing Time (16M)} & (e) \mbox{ CPU Comparisons (16M)} & (f) \mbox{ I/O (16M)}\\
\end{array}$
\caption{Sensitivity on input error for Hybrid-Hash algorithms}\label{fig:hhinput}
\label{fig:models}
\end{figure*}

For the hybrid-hash algorithms, the solid part in Figure~\ref{fig:pipeline} includes the time for grace partition, and the time for processing the resident partition in memory, while the gray part represents the time for recursive aggregation of the spilled partitions. Blocking in the hybrid-hash algorithms occurs mainly due to the aggregation of the resident partition. For larger memory sizes, the resident partition is larger, so it takes more time to aggregate all records of the resident partition, resulting in slightly longer times to first result for the hybrid-hash algorithms. For very large memory (4G) there is no grace partitioning, and since all records are in memory, they need to be fully aggregated before the first result is produced; thus the time to first result is also the time when the full aggregation is completed.

For both the Sort-based and the Hash-Sort algorithms, the solid part includes the time for generating sorted runs plus the time for merging sorted runs until the final merging round. The gray part indicates the time for the last merging phase, where the aggregation results are produced progressively during merging. As the memory size increases, the time to first result for the Sort-based algorithm increases because the time for merging is longer. For very small memory (1M) the Hash-Sort algorithm experiences a longer blocking time because it uses both hashing and sorting, while the hashing does not collapse many records. As memory increases, the hashing becomes more effective in collapsing which reduces both the sorting and merging time. For very large memory (4G), the Hash-Sort aggregates all records in memory and thus the time to first result is also the time to full aggregation (similarly to the hybrid-hash algorithms). 


\subsection{Input Error Sensitivity of Hybrid Hash}\label{sec:eval:hhinput}

The performance of all hybrid-hash algorithms is closely related to the input key cardinality $G$.  Note that $G$ serves as an exploit input of the hybrid-hash algorithm, as it is used to compute the number of partitions $P$. 
In practice the input set is not known in advance, so we estimate $G$. Since such estimation may not be accurate, we also tested the performance of the hybrid-hash algorithms assuming that $G$ is over/under-estimated. Using the dataset with cardinality ratio 0.02\%, we ran experiments where $P$ was computed assuming various (incorrect) values for $G$. In particular, we varied $G$ from a far over-estimated ratio (4096 times the actual cardinality) to a quite underestimated ratio (1/4096 of the actual cardinality). Figure~\ref{fig:hhinput} shows the experimental results for two different memory budgets (4M and 16M). When the input parameter is correct (i.e., the ratio is 1), the first memory configuration causes spills whereas the second memory configuration can be processed purely in memory. We also depict the running time of the Hash-Sort algorithm for comparison (since Hash-Sort does not depend on the parameter $G$).

Our experiments show that both overestimation and underestimation can affect the performance of the hybrid-hash algorithms. Specifically, an overestimation will cause unnecessary grace partitioning, and will thus increase the total I/O cost. In the worst case all hybrid-hash algorithms do grace partitioning, causing slower running times than the Hash-Sort algorithm. An underestimation will falsely process the aggregation earlier, resulting in less collapsing in the hybrid-hash and further grace partitioning. 

More specifically, the results in Figure~\ref{fig:hhinput} show that the Shared Hashing algorithm and the Original Hybrid-Hash may fallback to the Hash-Sort algorithm if the partition size is underestimated and turns out to be too large. The Dynamic Destaging algorithm works well in the underestimation case, as it always uses at least 50\% of the available memory for partitioning. Among all hybrid-hash algorithms, Pre-Partitioning achieves better tolerance to the error in the grouping key cardinality $G$; this is due to its guarantee that the in-memory partition will be completely aggregated. Pre-Partitioning has more robust performance for underestimated cases since it can still guarantee the complete aggregation of the resident partition, and it can also gather some statistics while aggregating the resident partition. It can then use the obtained statistics to guide the recursive processing of the spilled partitions.

\subsection{Hash Implementation Issues}\label{sec:eval:hash}

During the implementation of the hash-based algorithms (all four hybrid-hash algorithms, and also the Hash-Sort algorithm) we faced several issues related to the proper usage of hashing. Considering the quality of the hash function, we used Murmur hashing \cite{bibd:url///murmurhash}. We tried the multiplication method \cite{DBLP:books/aw/Knuth73} (the default hashing strategy in Java) in our experiments, but we found that its hash collision behavior deteriorated greatly for the larger grouping key cardinalities in our test datasets. Another issue related to the usage of the notion of hash function family for the hybrid-hash algorithms. It is important to have non-correlated hash functions for the two adjacent hybrid-hash levels. In our experiments we used Murmur hashing with different seeds for the different hybrid-hash levels. 

\begin{figure}[h]
\centering$
\begin{array}{cccc}
\psfig{file=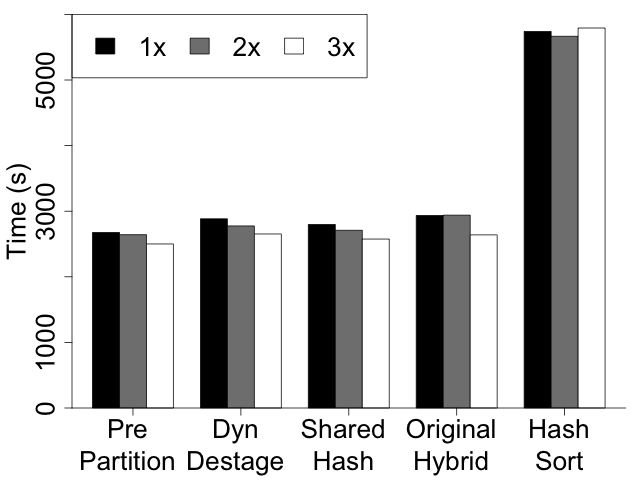, width=1.8in} &
\psfig{file=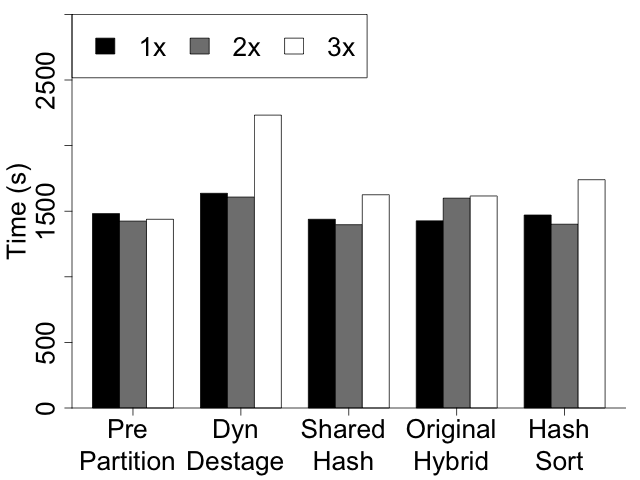, width=1.8in} \\
& \\
& \\
(a) \mbox{ Hash table size (2M)} & (b) \mbox{ Hash table size (4G)}
\end{array}$
\caption{Running time with different hash table sizes (as the ratios of number of slots over the hash table capacity).}\label{fig:htsize}
\end{figure}

We also examined how the hash table size (slot table size, or the number of slots in the slot table) affects performance. Given a fixed memory space, an in-memory hash table with a larger number of slots (which could potentially reduce hash collisions) in its slot table would have a smaller list storage area (so a smaller hash table capacity). Thus, the number of slots should be properly picked to trade-off between the number of hash collisions and the hash table capacity. In literature, it is often suggested to use a slot table size that is around twice the number of unique groups that can be maintained in the list storage area. Figure~\ref{fig:htsize} depicts the running times of the hash-based algorithms with varying slot table sizes (set to be 1x, 2x and 3x the number of unique groups maintained). In the small memory case, different slot table sizes do not affect the total running time significantly. In the larger memory case, all hash-based algorithms can aggregate the data in-memory when the slot table size is 1x (equal to the number of unique keys). Most algorithms do in-memory aggregation except for the Original Hybrid-Hash, which spills due to the larger slot table overhead. When the slot table size is 3x, only the Pre-Partitioning algorithm can complete the aggregation in-memory, because it always fills up the memory for resident partition before trying to spill. (In all other experiments we picked 1x so that all hash-based algorithms can finish in-memory for 4G memory).

\begin{figure}[h]
\centering$
\begin{array}{cccc}
\psfig{file=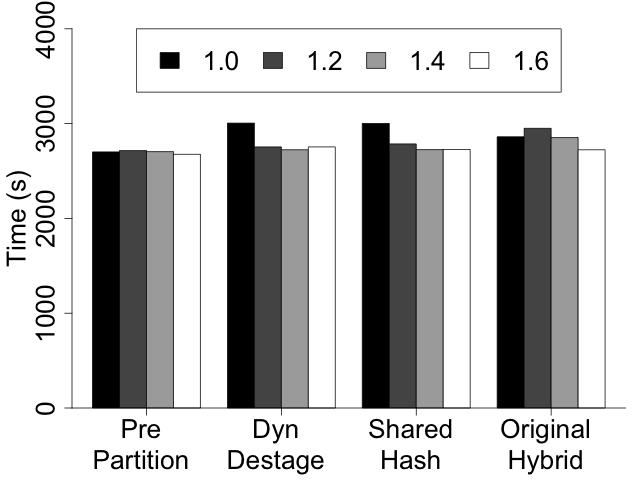, width=1.8in} &
\psfig{file=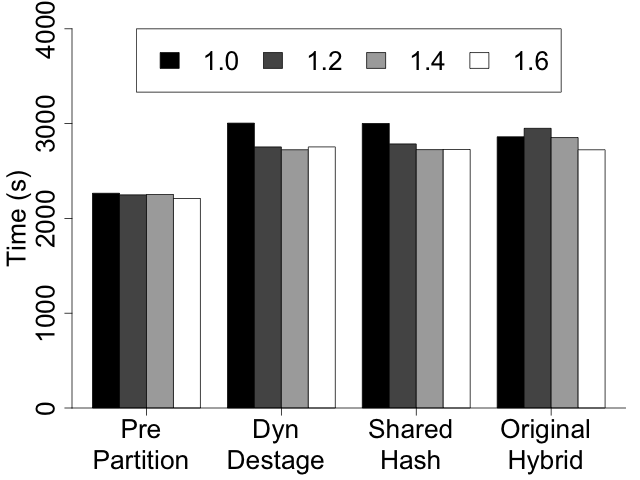, width=1.8in}\\
& \\
& \\
(a) \mbox{ Fudge factor (2M)} & (b) \mbox{ Fudge factor (4G)}
\end{array}$
\caption{Running time with different fudge factors.}\label{fig:fudge}
\end{figure}

Finally, we also explored the importance of the fudge factor $F$ in the hybrid-hash algorithms. This factor accounts for the extra memory overhead including both {\bf hash table overhead} (denoted as $o$) caused by the slot table and the list data structure other than the data itself, as well as {\bf extra overhead} (denoted as $f$) because of possible inaccurate estimations of the record size and memory page fragmentation. Here we define the fudge factor as $F = o * f$. Past literature has set the fudge factor to $1.2$, but it is not clear whether they have considered both kinds of overhead. In our experiments, the hash table overhead can be precisely computed based on the slot table structure; since we are using a linked-list-based table structure, there are 8 bytes of overhead for each slot table entry and 8 bytes of cost for each group in the list storage area. For the extra overhead, we tried four different ratios: $1.0$, $1.2$, $1.4$ and $1.6$. Figure~\ref{fig:fudge} shows the running times. We can see that clearly it is not wise to consider only the slot table overhead ($f = 1.0$) since the running times of the Dynamic Destaging and Shared Hashing algorithms increase in both memory configurations. This is because the smaller fudge factor causes an underestimated partition size $P$, and thus there are partitions that fail to be fit into the memory during the hybrid hash. From our experiments we also observed that using slightly larger $f$ values ($> 1.2$) has no significant influence on performance.

\section{Algorithm Selection}\label{sec:opt}

\vspace{-3mm}
\begin{figure}[h]
\begin{tabular}{cc}
\psfig{file=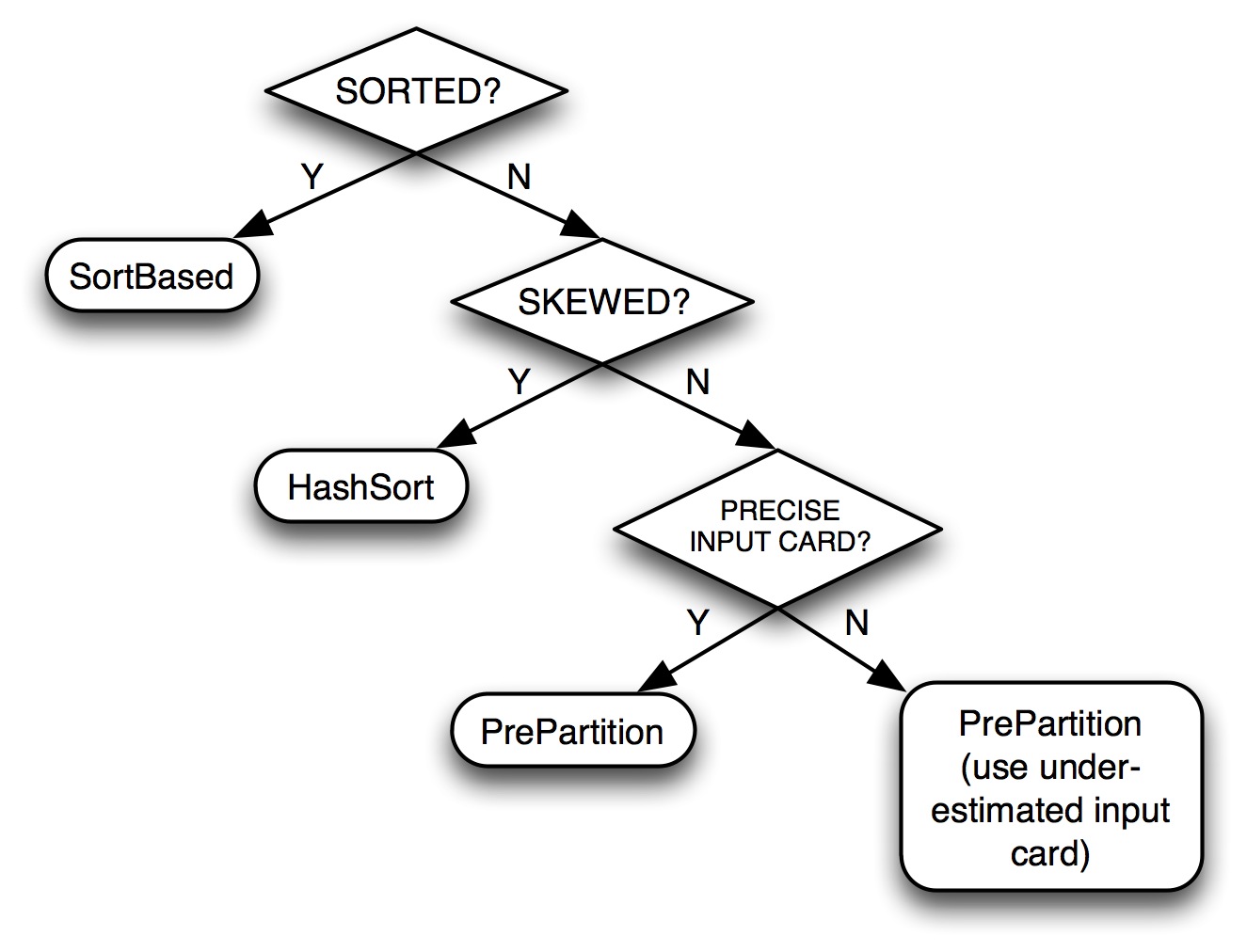, width=2.8in}
\end{tabular}
\caption{The decision tree for selecting the aggregation algorithm}\label{fig:decoct}
\vskip -6pt
\end{figure}

The observations from the experimental results in the previous section indicate that none of the algorithms can alone be the winner of all different cases. However, the Original Hybrid-Hash, Dynamic Destaging and Shared Hashing algorithms lose in most of the experiments compared with Pre-Partitioning, and their implementations are also rather complex. Thus, the final candidate algorithms from our experiments for AsterixDB will be the Sort-based, Hash-Sort and Pre-Partitioning algorithms. To choose the right algorithm from the three candidates, we can use the strategy as shown in Figure~\ref{fig:decoct} based on our observations. In detail:


\begin{itemize}
\item {\em When the input data is sorted:} the Sort-based algorithm can utilize the sorted order and compute the running aggregation through one scan. The other two algorithms need further I/O since the sorted property has no benefit for hashing.
\item {\em When the input data is skewed compared with the uniform-distributed dataset:} from the experiments in Section~\ref{sec:eval:skew}, we have seen that the Hash-Sort algorithm performs better than the others when the data is skewed. 
\item {\em When the input key cardinality is uncertain:} The Pre-Partitioning algorithm should be chosen for the uniform dataset when the key cardinality is uncertain. If the key cardinality is not precise, especially when it may be over-estimated, Pre-Partitioning may cause unnecessary grace partitioning with extra I/O cost. In practice, if the input key cardinality is not precise, the Pre-Partitioning algorithm can be used with an underestimated input key cardinality. This will force the Pre-Partitioning to do a hybrid-hash phase, and during this phase the algorithm can collect statistical information to adjust the input cardinality.
\end{itemize}

\section{Conclusions}\label{sec:conclu}

In this paper we have discussed our experiences when implementing efficient local aggregation algorithms for Big Data processing. We revisited the implementation details of six aggregation algorithms assuming a strictly bounded memory, and we explored their performance through precise cost models and extensive empirical experiments. Among the six aggregation algorithms, we proposed two new algorithm variants, the Hash-Sort algorithm and the Pre-Partitioning algorithm. In most cases, the four hybrid-hash algorithms were the preferred choice for better running time performance. The discussion in this paper guided our selection of the local aggregation algorithms in the recent release of AsterixDB \cite{DBLP:journals/pvldb/AlsubaieeAABBBCGHKLOPVW12}: the Pre-Partitioning algorithm for its tolerance on the estimation of the input grouping key cardinality, the Sort-based algorithm for its good performance when aggregating sorted data, and the Hash-Sort algorithm for its tolerance for data skew.
We hope that our experience can also help developers of other Big Data platforms to build the solid local aggregation fundamental. In AsterixDB, based on this work, we are now continuing our study of efficient aggregation implementations in a clustered environment, where more factors like per-machine workload balancing and network costs must be further considered. 

\begin{acknowledgements}
This work has been partially supported by NSF IIS awards 0910989 and 0910859, a grant from the UC Discovery program with a matching donation from eBay, and gifts from Google, hTC, Microsoft, and Oracle Labs.
\end{acknowledgements}

\bibliographystyle{spmpsci}      

\bibliography{biblibrary_jw}

\appendix\label{apedx}

\section{APPENDIX: Basic Component Models}

This section describes the details of the basic component models used in the cost model analysis. We use the symbols shown in Table~\ref{tbl:csymbols}.

\begin{center}
\vskip -12pt
\begin{table}[h]
\begin{tabular}{ll}
 Symbol                   &  Description                                \\[1ex]
\hline
$n$ & Number of raw records\\[1ex]
$m$ & Number of unique groups\\[1ex]
$A$ & A set of run files $\{A[1], ..., A[|A|]\}$\\[1ex]
$\mathcal{D}(n, m)$ & An input dataset of $n$ records and $m$ unique\\[1ex]
& groups\\[1ex]
$H$ & Hash table slots count\\[1ex]
$K$ & Hash table capacity in number of unique groups\\[1ex]
$M$ & Memory capacity in frames\\[1ex]
$\mathcal{U}$ & Number of unique keys, so that the dataset \\[1ex]
& $\mathcal{D}(n, m)$ can be generated through with- \\[1ex]
& replacement draws from this key set.
\end{tabular}
\caption{Symbols For Input Parameters}\label{tbl:csymbols}
\end{table}
\vskip -42pt
\end{center}

\subsection{Input Component}

There are two important quantities we will use in the algorithms' cost models. (1) Due to a restricted memory budget, a dataset $\mathcal{D}(n, m)$ will be processed in `chunks'. When considering a chunk of $r$ records ($r \leq n$), an important quantity is 
the number of unique keys that this chunk contains - denoted as $I_{key}(r, n, m)$ - assuming that the records are randomly picked from $\mathcal{D}(n, m)$. 
(2) Given a memory budget for $k$ groups (records of the form (\verb|key|, \verb|aggregated value|)), another important quantity is the number
of records - denoted as $I_{raw}\allowbreak(k, n, m)$ - that we should pick randomly from $\mathcal{D}(n, m)$ in order to fill up the memory with $k$ unique keys ($k \leq m$). Assuming draws without replacement, both quantities can be computed through direct application of Yao's formula \cite{DBLP:conf/icdt/GardyN99}. In particular:

\begin{align}
I_{key}(r, n, m) = & m * (1 - (1 - {{r}\over{n}})^{{n}\over{m}}) \label{equ:key_dep}\\
I_{raw}(k, n, m) = & n * (1 - (1 - {{k}\over{m}})^{{m}\over{n}}) \label{equ:raw_dep}
\end{align}






\subsection{Sort Component}
\label{sec-6-2}

When sorting the dataset $\mathcal{D}(n, m)$, we assume a 3-way-partition-quicksort \cite{DBLP:journals/siamcomp/Sedgewick77}. The required number of comparisons $C_{sort}\allowbreak(n, m)$ can be computed through a divide-and-conquer procedure by randomly choosing a split key and recursively sorting on the two sub-partitions:
\vspace{-2mm}
\begin{align}
C_{sort}(n, m) =& {\frac{n}{m}} * m - 1 + {1\over{m}}\sum_{i = 1}^{m}(C_{sort}(\frac{n}{m} * (m - i), m - i) \nonumber \\
&+ C_{sort}(\frac{n}{m} * (i - 1), i - 1))
\end{align}








Solving this recurrence we get the following formula:

\begin{equation}\label{equ:sort}
C_{sort}(n, m) = 2\frac{n}{m}(m - 1)ln(m - 2) + (\frac{n}{m} - 1)(2m - 3)
\end{equation}

\subsection{Merge Component}\label{theo:comp:merge}
\label{sec-6-3}

Consider a collection $A$ of sorted run files. Let $A[i]$ denote the size of the $i$-th file. Algorithm \ref{alg:merge} computes the cost for merging the collection $A$ using $M$ input buffer frames and the loser-tree based merging  method \cite{DBLP:books/aw/Knuth73}. By setting the cost function $F(A') (A' \subseteq A)$ to be the CPU comparisons in merging ($F(A')=log_{2}(|A'|)$) or the flushing I/O in merging ($F(A')=\sum_{i=1}^{|A'|}A'[i]$), the same algorithm can be used for either CPU comparison cost or the I/O cost. 

%

\begin{algorithm}
\caption{Algorithm for Merge Cost}\label{alg:merge}
\begin{algorithmic}
\REQUIRE $A$: files to be merged; $M$: available memory in frames; $F$: cost function.
\WHILE{$|A|>1$}
\STATE {\bf if} $|A| \leq M$:  all files can be merged in a single round. Add $F(A)$ to cost, and stop.
\STATE {\bf if} $M < |A| < 2M$: merge the first $(|A| - M + 1)$ files to produce a single run; remove the merged files from $A$ and add the new run at the end of $A$ ($|A|$ is thus reduced to $M$ files). Add $F(\{A[1], ..., A[|A| - M + 1]\})$ to $C_{merge}(A, M)$
\STATE {\bf if} $|A| \geq 2M$: merge the first $M$ files into a new run. Remove the merged files from $A$ and add the new run at the end of $A$. 
Add $F(\{A[1], ..., A[M]\})$ to $C_{merge}(A, M)$
\ENDWHILE
\end{algorithmic}
\end{algorithm}




\subsection{Hash Component}
\label{sec-6-4}

Consider a hash table with $H$ slots in the slot table; Let $K$ denote the maximum number of group records that can be stored in the list storage area. Note that duplicates are aggregated within group records, so filling up the list storage area would imply encountering $K$ unique groups. The number of records drawn randomly from $\mathcal{D}(n, m)$ to fill up the list storage area (i.e., to get $K$ unique keys) is thus $I_{raw}(K, n, m)$.

Let $C_{hash}\allowbreak(n, m, K, H)$ denote the number of comparisons needed to fill up the list storage area. This accounts for both hash hits (denoted as $c_{succ}$; these are records that have been seen already and are thus aggregated) and hash misses (denoted by $c_{unsucc}$; these are records that have not been seen before). 
For the $i$-th insertion to be a hash hit, it must correspond to a key which has already been inserted in the hash table. Using Equation \ref{equ:key_dep}, at the $i$-th insertion the number of unique keys already in the hash table is 
\[
k_{i} = I_{key}(i, n, m)
\]
Note that the no-replacement assumption of Yao's formula implies that after each insertion, the distribution of the remaining keys in the input set changes; this distribution is thus difficult to re-estimate after each insertion. Instead, we will assume here that the dataset $\mathcal{D}(n, m)$ is generated by randomly drawing keys {\bf with replacement} from a `generator' set with $\mathcal{U}$ unique keys. Then each insertion can be considered as a random pick from the $\mathcal{U}$ unique keys with replacement, and the probability for a hash hit for the $i$-th insertion becomes: 
\[
Pr_{hashHit} = {{k_{i}}\over{\mathcal{U}}}
\]

We note that the average number of unique keys $\tilde {m}$ in $n$ random draws is given by:

\begin{equation}\label{equ:gen_key}
\tilde{m} = \mathcal{U} * (1 - (1 - {{1}\over{\mathcal{U}}})^{n})
\end{equation}

We can then estimate $\mathcal{U}$ by substituting the expected value $\tilde{m}$ with $m$ in the above equation. 

To compute the number of comparisons during a hash hit we need the expected number of groups contained in a non-empty slot, assuming the probability of finding a match at any group along the slot's linked list is the same. The number of non-empty slots in the hash table at the $i$-th insertion can be calculated using the urn model \cite{DBLP:books/crc/tucker97/MotwaniR97} as 
\begin{equation}\label{equ:uslots}
H_{u}(i, H, n, m) = H * (1 - (1 - {1\over{H}})^{k_{i}})
\end{equation}
Then the expected number of groups in a non-empty slot would be
\[
L_{slot} = {{k_{i}}\over{H_{u}(i, H, n, m)}}
\]
The expected comparison cost for a hash hit becomes:

\begin{equation}
c_{succ}(i, n, m, H)  = Pr_{hashHit} * {{L_{slot} + 1}\over{2}}
\end{equation}

A hash miss happens when a record is hashed either into a previously empty slot (this does not require a comparison) or into a non-empty slot but where no match is found (this case will incur comparisons until the end of the linked list is reached). The probability that it is inserted into a non-empty slot is:
\[
Pr_{nonempty} = {{H_{u}(i, H, n, m)}\over{H}}
\]
and thus the hash miss comparison cost then becomes:

\begin{align}
c_{unsucc}(i, n, m, H) =&  (1 - Pr_{hashHit})  * L_{slot} \nonumber \\
&* Pr_{nonempty}
\end{align}

Finally, the total comparison cost is given by:

\begin{align}\label{equ:hash}
C_{hash}(n, m, K, H) &=& \sum_{i = 0}^{I_{raw}(K, n, m)} (c_{succ}(i, n, m, H) + \nonumber \\
&& c_{unsucc}(i, n, m, H))
\end{align}

For some hybrid-hash algorithms a spilled partition may contain both aggregated groups and non-aggregated  records. To insert such a ``mixed'' dataset into a hash table, the cost model should be adjusted. Let $u$ denote the number of aggregated groups (which are thus unique) and $n$ the number of `raw' (not yet aggregated) records. 
To insert the $u$ unique groups the comparisons arise only from hash misses:

\begin{align}
c_{unique}(n, m, H, u) &=& \sum_{i = 1}^{u}({{H_{u}(i - 1, H, n, m)}\over{H}} \nonumber \\
&&* {{i - 1}\over{H_{u}(i - 1, H, n, m)}})
\end{align}

For calculating the number of comparisons from the insertion of the raw records after inserting the $u$ unique groups, we first note that the probability for a hash hit is:
\[
Pr'_{hashHit} = {{k_{i} + u}\over{\mathcal{U}}}
\]

The expected number of groups in a non-empty slot for the $i$-th insertion is given by:
\[
L'_{slot} = {{k_{i} + u}\over{H_{u}(i, H, n, m)}}
\]
So the hash comparison cost if the $i$-th insertion is a hash hit is:

\begin{equation}
c_{succ}(i, n, m, H, u)  = Pr'_{hashHit} * {{L'_{slot} + 1}\over{2}}
\end{equation}

To calculate the hash miss cost, we note that the probability that the insertion is to a non-empty slot is adjusted as:
\[
Pr'_{nonempty} = {{H_{u}(i + u, H, n, m)}\over{H}}
\]

Hence the comparison cost for the hash miss of the $i$-th insertion becomes: 

\begin{equation}
c_{unsucc}(i, n, m, H, u) = L'_{slot} * Pr'_{nonempty} * (1 - Pr'_{hashHit})  
\end{equation}
The overall cost for the `mixed' input case, denoted by $C_{hash}\allowbreak(n, m, K, H, u)$ is thus:

\begin{align}\label{equ:hash_preinsert}
C_{hash}(n, m, K, H, u) =& c_{unique}(n, m, H, u) \nonumber \\
&+ \sum_{i = 1}^{I_{raw}(K - u, n, m)} (c_{succ}(i, n, m, H, u) \nonumber \\
& + c_{unsucc}(i, n, m, H, u))
\end{align}







\end{document}